\documentclass{PoS}
\usepackage[sumlimits,intlimits,namelimits]{amsmath}
\usepackage{mathtools}
\usepackage{graphicx}
\usepackage{grffile}
\usepackage[noabbrev]{cleveref}
\usepackage{xcolor}
\usepackage{subfig}
\usepackage{macros}
\usepackage{bm}
\usepackage[backend=biber,
backref=false,
style=numeric-comp,
sorting=none,
block=ragged,
firstinits=true,bibencoding=utf8
]{biblatex}

\renewbibmacro*{pageref}{}
\renewbibmacro*{in:}{\printtext{\vphantom{a}}}
\DeclareFieldFormat{pages}{#1}
\DefineBibliographyStrings{english}{mathesis = {Master's thesis}}
\providecommand{\texorpdfstring}[2]{#1}
\usepackage[%
disable
]{todonotes}

\newcommand{\sub}[1]{_\mathrm{#1}}
\newcommand{\abs}[1]{|#1|}

\title{Light-Quark Resonances at COMPASS}

\ShortTitle{Light-Quark Resonances at COMPASS}

\author{\speaker{Stefan Wallner}\thanks{For the COMPASS collaboration}\\
	Institute for Hadronic Structure and Fundamental Symmetries (TU Munich)\\
	E-mail: \email{stefan.wallner@tum.de}}

\abstract{%
	The main goal of the spectroscopy program at COMPASS is to explore the light-meson spectrum in the mass range below about $2\,\text{GeV}/c^2$ using diffractive dissociation reactions. Our flagship channel is the production of three charged pions in the reaction: $\pi^- + p \to \pi^-\pi^-\pi^+ + p_\text{recoil}$, for which COMPASS has acquired the so far world's largest dataset of roughly $50\,\text{M}$ exclusive events using an $190\,\text{GeV}/c$  $\pi^-$ beam.
	Based on this dataset, we performed an extensive partial-wave analysis. 
	
	In order to extract the parameters of the $\pi_J$ and $a_J$ resonances that appear in the $\pi^-\pi^-\pi^+$ system, we performed the so far most comprehensive resonance-model fit, using Breit-Wigner parametrizations.
	This method in combination with the high statistical precision of our data allows us to study ground and excited states.
	We study the $a_4(2040)$ resonance in the $\rho(770)\pi G$ and $f_2(1270)\pi F$ decays.
	In addition to the ground state resonance $a_1(1260)$, we have found evidence for the $a_1(1640)$, which is the first excitations of the $a_1(1260)$, in our data.
	We also study the spectrum of $\pi_2$ states by simultaneously describing four $J^{PC}=2^{-+}$ waves using three $\pi_2$ resonances, the $\pi_2(1670)$, the $\pi_2(1880)$, and the $\pi_2(2005)$.
	
	Using a novel analysis approach, where the resonance-model fit is performed simultaneously in narrow bins of the squared four-momentum transfer $t'$ between the beam pion and the target proton, allows us to study the $t'$ dependence of resonant and non-resonant components included in our model.
	We observe that for most of the partial waves, the non-resonant components show a steeper $t'$ spectrum compared to the resonances and that the $t'$ spectrum of most of the resonances becomes shallower with increasing resonance mass.
	We also study the $t'$ dependence of the relative phases between resonance components. The pattern we observe is consistent with a common production mechanism of these states.
}

\FullConference{XIII Quark Confinement and the Hadron Spectrum - Confinement2018\\
		31 July - 6 August 2018\\
		Maynooth University, Ireland}

\makeatletter
\let\@fnsymbol\@alph
\makeatother

\newcommand{\twoPlotWidth}{0.441\linewidth}
\newcommand{\threePlotWidth}{0.32\linewidth}
\begin{document}
\todo[inline]{references
	
	[1,2,5-8] remove line break after author

	Removed arXiv links??}
	\section{Light-meson spectroscopy at COMPASS}
	COMPASS is a fixed-target multi-purpose experiment located at CERN.
	Positive and negative secondary hadron beams or a tertiary muon beam are directed onto various types of targets.
	The forward-going final-state particles are detected by a two-stage magnetic spectrometer, which has a large acceptance over a broad kinematic range.
	
	The main goal of the COMPASS spectroscopy program is to study the light-meson spectrum up to masses of about \SI{2}{\GeVcc}.
	At COMPASS, these light mesons are produced in diffractive scattering of the \SI{190}{\GeVc} negative hadron beam, which contains mainly negative pions, off a liquid-hydrogen target. In these reactions, we can study $a_J$- and $\pi_J$-like mesons.
	Due to their very short lifetime, we observe the resonances only in their decays into quasi-stable final-state particles. Our flagship channel is the decay into three charged pions: \reaction.
	COMPASS has acquired the so far world's largest dataset of about \SI{50}{M} exclusive events for this channel, which allows us to apply novel analysis methods~\cite{Adolph2015}.
	
	\section{Analysis method}
	\label{sec:method}
	
	\subsection{Partial-wave decomposition}
	\label{sec:method:pwd}
	We employ the method of partial-wave analysis in a two-step approach.
	In the first step, called partial-wave decomposition, data are decomposed into contributions from various partial waves~\cite{Adolph2015}.
	To this end, we construct a model for the intensity distribution $\mathcal{I}(\tau)$ of the \threePi final state in terms of the five-dimensional phase-space variables of the $3\pi$ system that are represented by $\tau$. Using the isobar approach,  $\mathcal{I}(\tau)$ is modeled as a coherent sum of partial-wave amplitudes, which are defined by the quantum numbers of the $3\pi$ system ($\JPCMrefl$),\footnote{Here, $J$ is the spin of the $3\pi$ state, $P$ its parity and $C$ its charge conjugation quantum number. The spin projection of $J$ along the beam axis is given by $M^\varepsilon$.} the intermediate \twoPi resonance $\zeta$ through which the decay proceeds, and the orbital angular momentum $L$ between the bachelor pion and the isobar. These quantum numbers are represented by $a = \Wave{J}{P}{C}{M}{\varepsilon}{\zeta}{\PpiNeg}{L}$:
	\begin{equation}
		\mathcal{I}(\tau; \mThreePi, \tpr) = \left|\sum\limits_{a}^{\text{waves}} {\mathcal T_a(\mThreePi, \tpr)}\, {\psi_a(\tau;\mThreePi, \tpr)}\right|^2.
	\end{equation}
	Within the isobar model, the decay amplitudes $\psi_a$ can be calculated. This allows us to extract the partial-wave amplitudes $\mathcal T_a$, which determine the strength and phase of each wave, from the data by an unbinned maximum-likelihood fit.
	
	In order to extract the \mThreePi dependence of the partial-wave amplitudes, the maximum-likelihood fit is performed independently in narrow bins of the three-pion mass \mThreePi.
	In addition, the large size of the dataset allows us to study the dependence of the partial-wave amplitudes on the squared four-momentum transfer \tpr, by binning the data in \tpr as well.
	By performing the partial-wave decomposition independently in $100$ \mThreePi bins in the range $0.5 < \mThreePi < \SI{2.5}{\GeVcc}$ and $11$ \tpr bins in the range $0.1 < \tpr < \SI{1.0}{\GeVcsq}$,\footnote{By definition, \tpr is a positive quantity, as $\tpr\equiv \abs{t} - \abs{t}\sub{min}$.} we extract simultaneously the \mThreePi and \tpr dependence of the partial-wave amplitudes from the data.
	
	We use a wave-set of 88 partial waves with spin $J$ and angular momentum $L$ up to six, which includes six different isobars. This is the largest wave-set used so far for this channel.
	\subsection{Resonance-model fit}
	\label{sec:method:rmf}
	In the second step of this analysis, the so-called resonance-model fit, we parameterize the \mThreePi dependence of the partial-wave amplitudes in order to extract the masses and widths of resonances appearing in the \threePi system. Details can be found in ref.~\cite{Adolph2018}.
	We model the amplitude for a wave $a$ as a coherent sum over wave components $j$ that we assume to contribute to this wave:\footnote{For simplicity, we absorb here the $\sqrt{\mThreePi}$ factor, the phase space integral, and the production factor in the dynamical amplitudes (see ref.~\cite{Adolph2018} for details).}
	\begin{equation}
		\mathcal{T}_a(\mThreePi, \tpr) =  \smashoperator[r]{\sum\limits_{j\in\mathbb{S}_a}} \mathcal{T}_a^{j}(\mThreePi, \tpr)= \smashoperator[r]{\sum\limits_{j\in \mathbb{S}_a}}\mathcal{C}^{j}_a(\tpr)\cdot \mathcal{D}^{j}(\mThreePi, \tpr; \zeta_j). \label{eq:rm:transition-amplitude}
	\end{equation}
	The \textit{dynamical amplitudes} $\mathcal{D}^j(\mThreePi, \tpr; \zeta_k)$ represent the resonant and non-resonant wave  components. We use relativistic Breit-Wigner amplitudes for resonances and a phenomenological parameterization for the non-resonant components.\footnote{The leading \mThreePi dependence of the non-resonant term is $e^{-c \tilde{q}^2(\mThreePi)}$, where $\tilde{q}(\mThreePi)$ is an approximation for the two-body break-up momentum of the isobar pion system, which is also valid below threshold.}
	Each dynamical amplitude is multiplied by a \textit{coupling amplitude} $\mathcal{C}^{j}_a(\tpr)$, which determines the strength and phase with which each wave component $j$ contributes to the corresponding wave $a$. We use independent coupling amplitudes for each \tpr bin. Thus, also in this analysis step, we do not impose a model for the \tpr dependence of the amplitudes of the wave components.
	We simultaneously describe all $11$ \tpr bins in a single resonance-model fit, keeping the mass and width parameters of each resonance component the same in each \tpr bin. In this way, our \tpr-resolved analysis gives us an additional dimension of information, which helps to better separate resonant from non-resonant components.
	Furthermore, it allows us to extract the \tpr dependence of each wave component in different partial waves as discussed below.
	
	Our two-step approach allows us to select a subset of waves to be included in the resonance-model fit. In this analysis, we select a subset of 14 out of the 88 partial waves with $\JPC = \zeroMP$, \onePP, \oneMP, \twoPP, \twoMP, and \fourPP quantum numbers. This subset accounts for about \SI{60}{\percent} of the total intensity. The 14 waves are parameterized by 11 resonances: \PpiEighteenhundret, \PaOne, \PaOneFourteenTwenty, \PaOnePr, \PpiOne, \PaTwo, \PaTwoPr, \PpiTwo, \PpiTwoPr, \PpiTwoPrPr, and \PaFour plus a non-resonant component in each wave.
	
	After separating the individual resonant and non-resonant components in the resonance-model fit, we can determine the \tpr dependence of their intensities, i.e. the \tpr spectra, and the \tpr dependence of the relative phases of the coupling amplitudes.
	Integrating the intensity over \mThreePi gives the \tpr-dependent yield, i.e. the \tpr spectrum, which is the number of events of wave component $j$ in wave $a$ in the $11$ \tpr bins:
	\begin{equation}
		\mathcal{I}^j_a(\tpr) = \frac{1}{\Delta\tpr}\int\limits_{m_\mathrm{min}}^{m_\mathrm{max}}\mathrm{d}\mThreePi\, \left|\mathcal{T}^j_a(\mThreePi, \tpr)\right|^2.
	\end{equation}
	To account for the non-equidistant \tpr-binning, we normalize in each \tpr bin the intensity to the respective bin width $\Delta\tpr$.
	The intensity of most wave components falls approximately exponentially with increasing \tpr. This is consistent with the expectation from Regge theory. For waves with spin projection $M \neq 0$, the intensity is kinematically suppressed for small values of \tpr by an additional $(\tpr)^{|M|}$ factor~\cite{Perl:1974}. Therefore, we fit the model
	\begin{equation}
		\mathcal{I}^j_a(\tpr) = A^j_a \cdot \left(\tpr\right)^{|M|}\cdot e^{-b^j_a\tpr} \label{eq:method:tspectrum}
	\end{equation}
	to the \tpr spectra extracted from the resonance-model fit. The real-valued parameter $A^j_a$ and the slope parameter $b^j_a$ are left free in the fit.
	Thereby, we extract one slop parameter for each wave component $j$ and for each partial wave $a$ that includes this wave component.
	
	Special cases are resonance components that appear in different partial waves with the same \JPCMrefl quantum numbers, but different decay modes $\zeta \Ppi L$.
	As the \tpr dependence is a property of the production and not of the decay, the same resonance component in different decay modes should show the same \tpr dependence. We incorporate this constraint into our model by fixing the \tpr dependence $\mathcal{C}^j_b(\tpr)$ of a resonance $j$ that appears in wave $b$ to the \tpr dependence $\mathcal{C}^j_a(\tpr)$ in a reference wave $a$ via
	\begin{equation}
		\mathcal{C}^j_b(\tpr) = \prescript{}{b}{\mathcal{B}}_a^j\,\mathcal{C}^j_a(\tpr).\label{eq:method:branching}
	\end{equation}
	The \tpr-independent complex-valued \textit{branching amplitude} $\prescript{}{b}{\mathcal{B}}_a^j$ represents the relative strength and phase between the two decay modes and is the only remaining free parameter for resonance component $j$ in wave $b$.
	By this constraint, the \tpr spectra of resonance $j$ in waves $a$ and $b$ are the same, except for a \tpr-independent scaling factor proportional to $|\prescript{}{b}{\mathcal{B}}_a^j|^2$.\footnote{In addition to $|\prescript{}{b}{\mathcal{B}}_a^j|^2$, the different phase-space integrals enter the scaling factor. Furthermore, the production amplitude causes a slight \tpr dependence of this factor, which we drop here. See ref.~\cite{Adolph2018} for details.}
	Also, the \tpr dependence of the phase of resonance $j$ in waves $a$ and $b$ is the same, except for the \tpr-independent phase offset $\mathrm{arg}[\prescript{}{b}{\mathcal{B}}_a^j]$.

	The fit result is affected by many systematic effects, e.g. the choice of the parameterizations of the wave components, the selected 14-wave sub-set, or the constraints imposed by \cref{eq:method:branching}. We performed more than 200 systematic studies to improve our model, to study the evidence for some resonance signals, and to determine the systematic uncertainties of the extracted parameters.
	
	\section{Selected Results of the resonance-model fit}
	In the following subsections, we present selected results for resonances with $\JPC=\fourPP$, \onePP, and \twoMP quantum numbers.
	As the statistical uncertainties are at least one order of magnitude smaller than the systematic ones, we quote only systematic uncertainties.
	A more exhaustive discussion of the results of the resonance-model fit can be found in ref.~\cite{Adolph2018}.
1	\subsection{$\bm{\JPC = 4^{++}}$ resonances}
	\begin{figure}[tbp]
		\subfloat[]{%
			\includegraphics[width=\threePlotWidth]{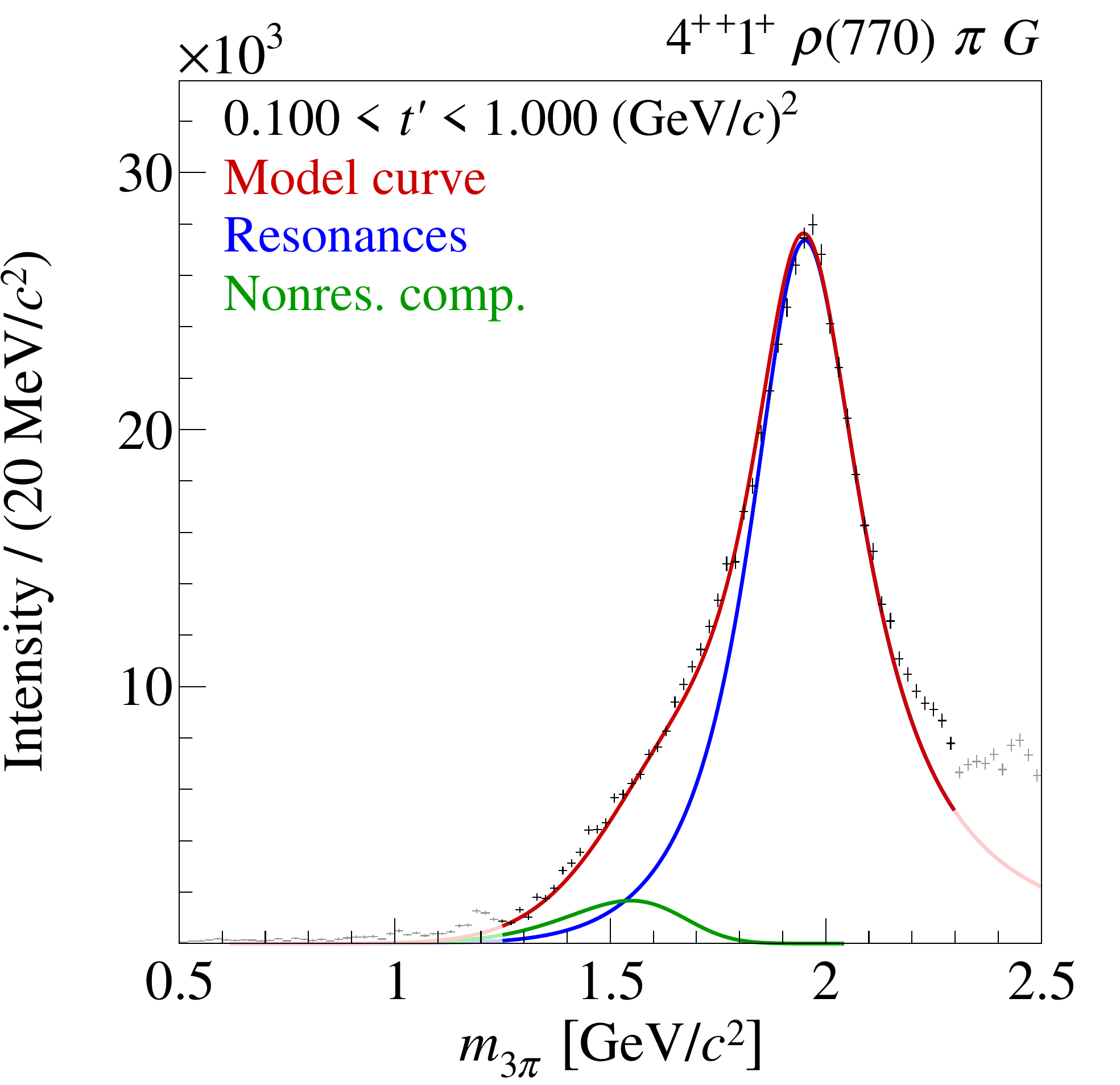}%
			\label{fit:res:4pp:rho}%
		}%
		\subfloat[]{%
			\includegraphics[width=\threePlotWidth]{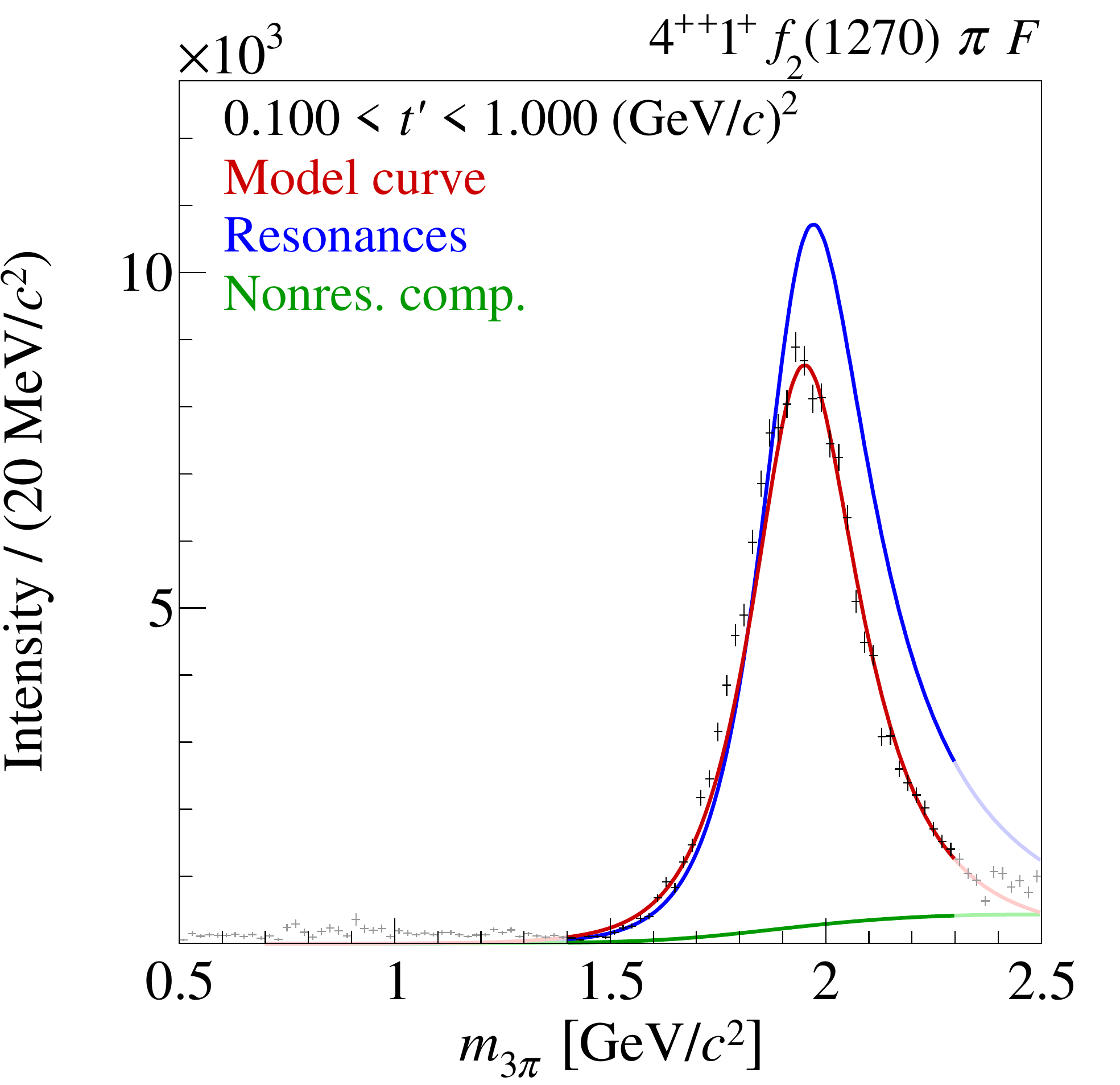}%
			\label{fit:res:4pp:f2}%
		}%
		\subfloat[]{%
			\includegraphics[width=\threePlotWidth]{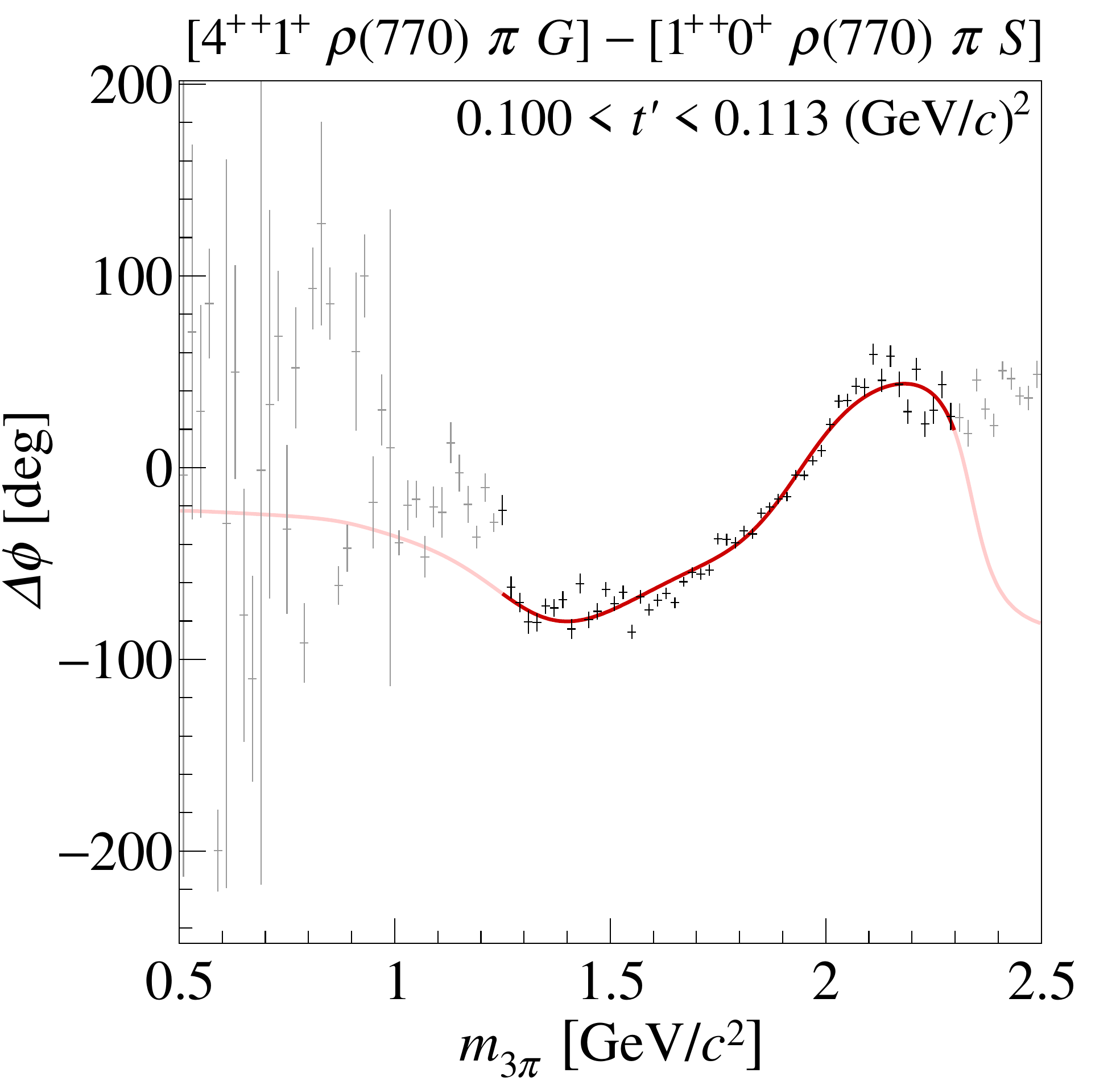}%
			\label{fit:res:4pp:f2phase}%
		}%
		\caption{\tpr-summed intensity distribution of (a) the \Wave 4++1+\Prho\Ppi G wave and (b) the \Wave 4++1+\PfTwo\Ppi F wave. (c) shows the relative phase of the \Wave 4++1+\Prho\Ppi G wave with respect to the \Wave 1++0+\Prho\Ppi S wave in the lowest \tpr bin. The data points show the result of the partial-wave decomposition. Uncertainties are statistical only. The red curves represent the resonance model. The blue curves represent the \PaFour resonance. The green curves represent the non-resonant components. The extrapolations beyond the fitted \mThreePi range are shown in lighter colors.}
		\label{fit:res:4pp}
	\end{figure}
	\begin{figure}[tbp]
		\centering
		\subfloat[]{%
			\includegraphics[width=\twoPlotWidth]{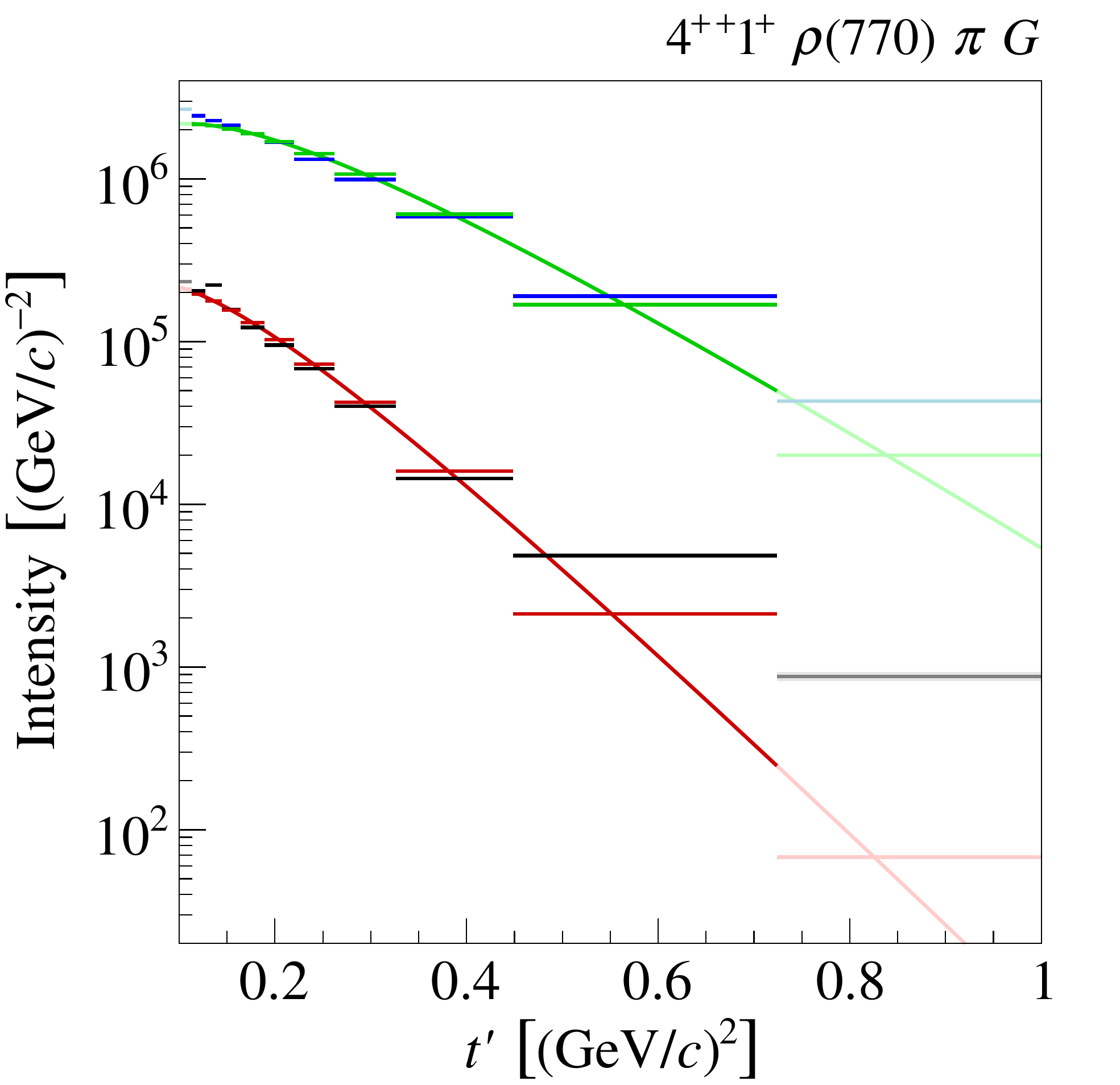}%
			\label{fit:res:4ppt:rho}%
		}%
		\subfloat[]{%
			\includegraphics[width=\twoPlotWidth]{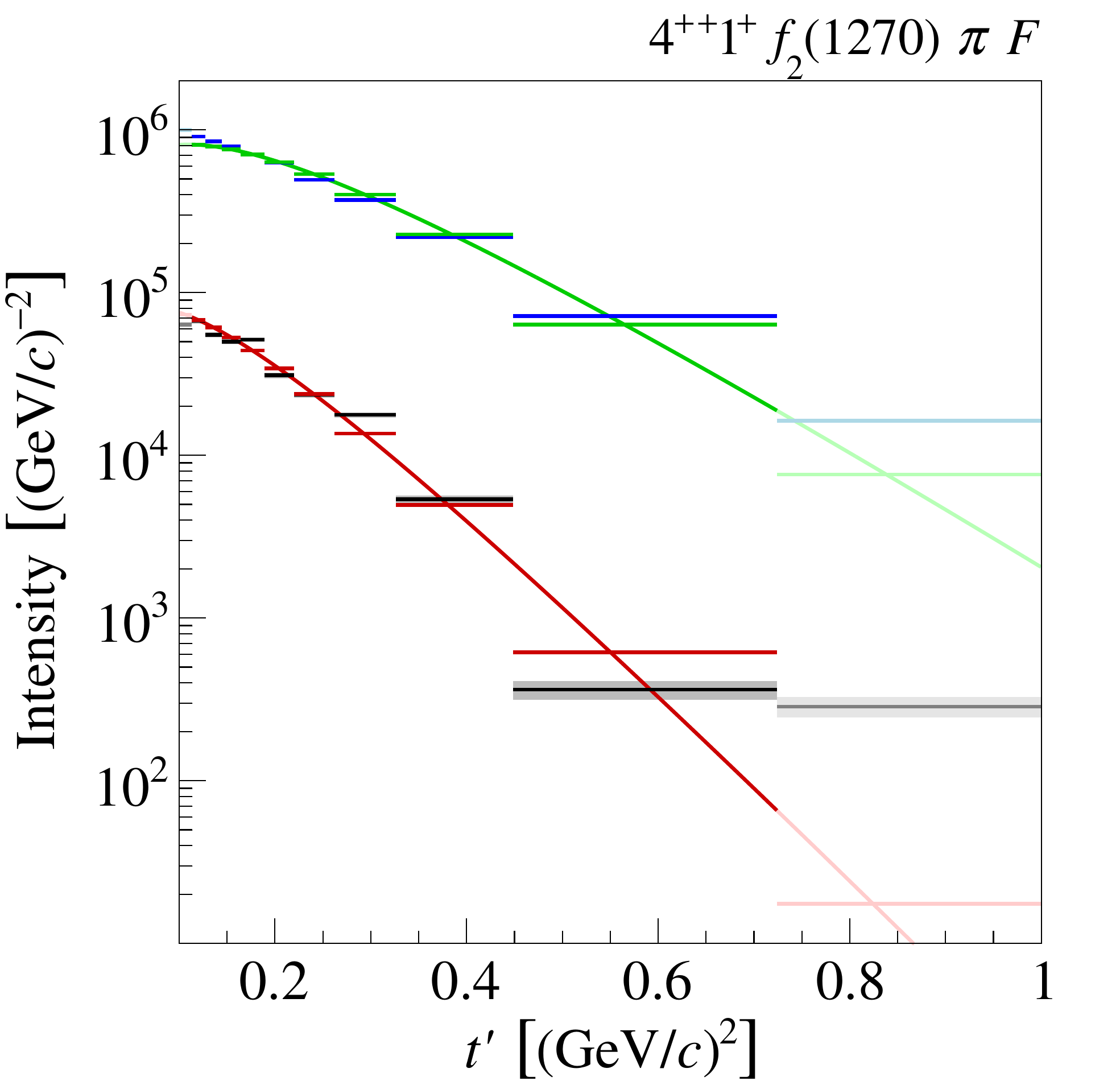}%
			\label{fit:res:4ppt:f2}%
		}%
		\caption{\tpr spectra of the \PaFour and the non-resonant components in (a) the \Wave 4++1+\Prho\Ppi G wave and (b) the \Wave 4++1+\PfTwo\Ppi F wave. The black horizontal lines indicate the central values and the gray boxes the statistical uncertainties of the \tpr spectra of the non-resonant component. The blue lines and boxes represent the \tpr spectra of the \PaFour. The red and green curves and lines show the results of a fit of \cref{eq:method:tspectrum} to these data.}
		\label{fit:res:4ppt}
	\end{figure}
	\Cref{fit:res:4pp:rho} shows the intensity distribution of the \Wave 4++1+\Prho\Ppi G wave summed over all 11 \tpr bins (\tpr-summed). It exhibits a clear peak at about \SI{2}{\GeVcc}, which is nearly completely described by the \PaFour resonance component. The relative phase of the  \Wave 4++1+\Prho\Ppi G wave with respect to, e.g., the \Wave 1++0+\Prho\Ppi S wave is shown in \cref{fit:res:4pp:f2phase}. We observe a clear rise in the \SI{2}{\GeVcc} mass region, as expected for a resonance. The low-mass tail of the peak is described as an interference effect between the \PaFour and the non-resonant component in this wave.
	Also in the \Wave 4++1+\PfTwo\Ppi F wave, we observe a clear \PaFour signal as shown in \cref{fit:res:4pp:f2}.
	Our estimates for the \PaFour resonance parameters are $m_0 = 1935^{+11}_{-13}\,\si\MeVcc$ and $\Gamma_0 = 333^{+16}_{-21}\,\si\MeVcc$. They are comparably robust with respect to systematic effects, due to the small non-resonant contributions in these \fourPP waves.
	Our estimates for the \PaFour mass and width agree with previous measurements, based on diffractive production~\cites{alekseev:2009aa}{Adolph:2014rpp}{chung:2002pu}{Amelin:1999gk}. However, our values is at variance with other measurements, e.g. from production in $\bar pp$ collisions~\cite{anisovich:2001pn}.
	
	Including both, the \Prho \Ppi and the \PfTwo\Ppi decay modes of the \PaFour in a resonance-model fit not only improves the accuracy of the estimated \PaFour parameters, but also allows to extract the branching-fraction ratio between these two decay modes. Correcting the measured yields for the unobserved decays to the \threePiN final state\footnote{We correct the yields for the unobserved decays $\PaFour*^- \to \Prho*^-\PpiNeg$ and $\PaFour*^- \to \PfTwo*^-\PpiNeg$ to the \threePiN final state assuming isospin symmetry. We also include the branching fraction of the \PfTwo into
		$2\pi$ and corrections of the isospin factor due to self-interference
		effects  (see ref.~\cite{Adolph2018} for details).} gives a branching-fraction ratio of $B^{\PaFour*, \mathrm{corr}}_{\Prho*\Ppi G, \PfTwo*\Ppi F} = 2.9^{+0.6}_{-0.4}$.
	
	The \tpr spectra of the \PaFour and the of non-resonant components in the \fourPP waves are shown in \cref{fit:res:4ppt}. The \tpr spectra of the \PaFour component in both waves are constrained by \cref{eq:method:branching} to have the same shape.
	The data are in fair agreement with the exponential model in \cref{eq:method:tspectrum}, except for the highest \tpr bin where the model systematically underestimates the data.
	Our estimate for the slope parameter of $b = 9.2^{+0.8}_{-0.5}\,\si\perGeVcsq$ lies within the range we observe for most of the resonances.
	The non-resonant components show a steeper falling \tpr spectrum with slopes of $b = 14\pm 4\,\si{\perGeVcsq}$ for the $\Prho\Ppi G$ decay and $b = 14.5^{+1.8}_{-3.7}\,\si\perGeVcsq$ for the $\PfTwo\Ppi F$ decay.
	
	\subsection{$\bm{\JPC = 1^{++}}$ resonances}
	\begin{figure}[tbp]
		\subfloat[]{%
			\includegraphics[width=\threePlotWidth]{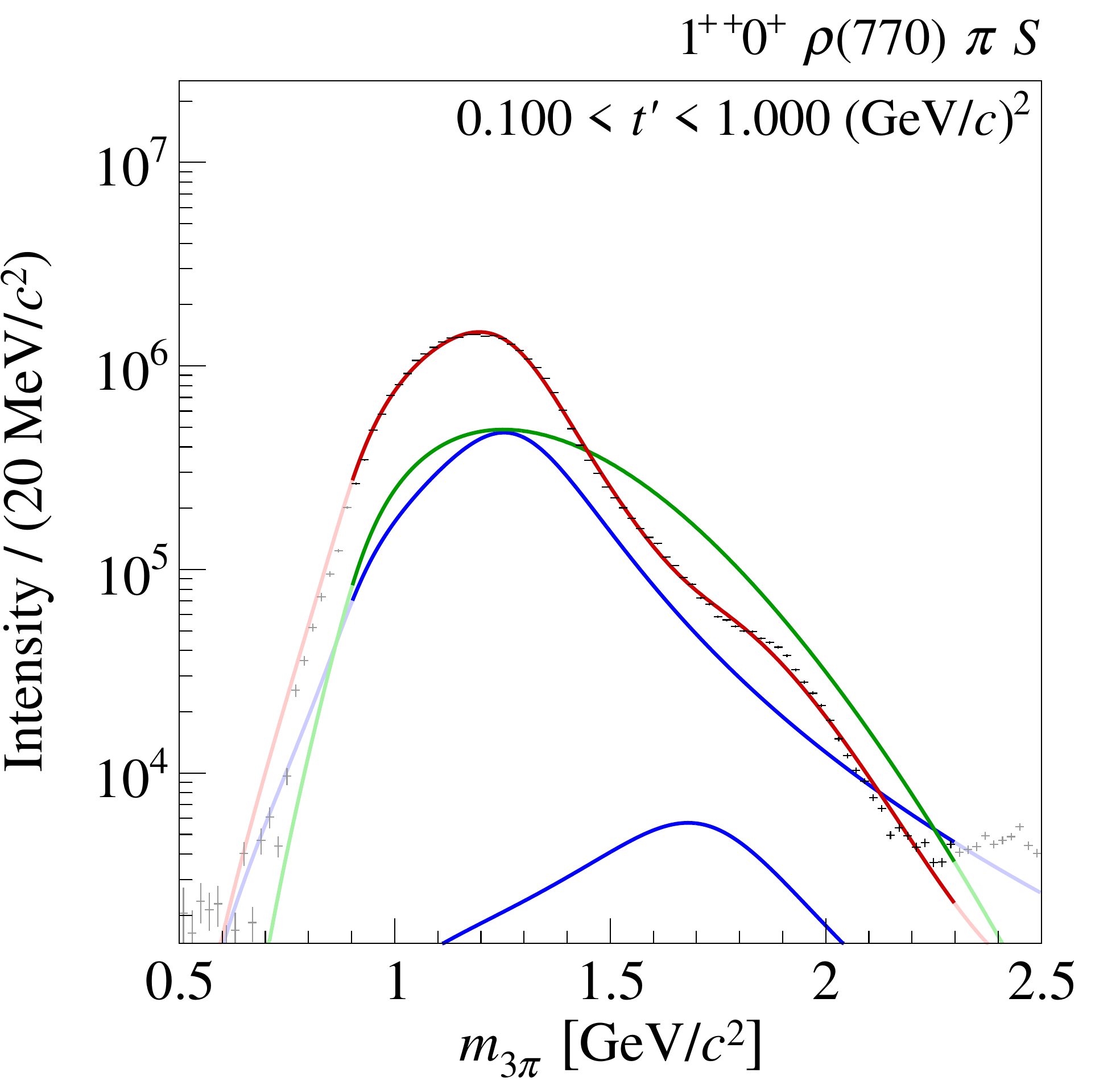}
			\label{fit:res:1pp:rho}%
		}%
		\subfloat[]{%
			\includegraphics[width=\threePlotWidth]{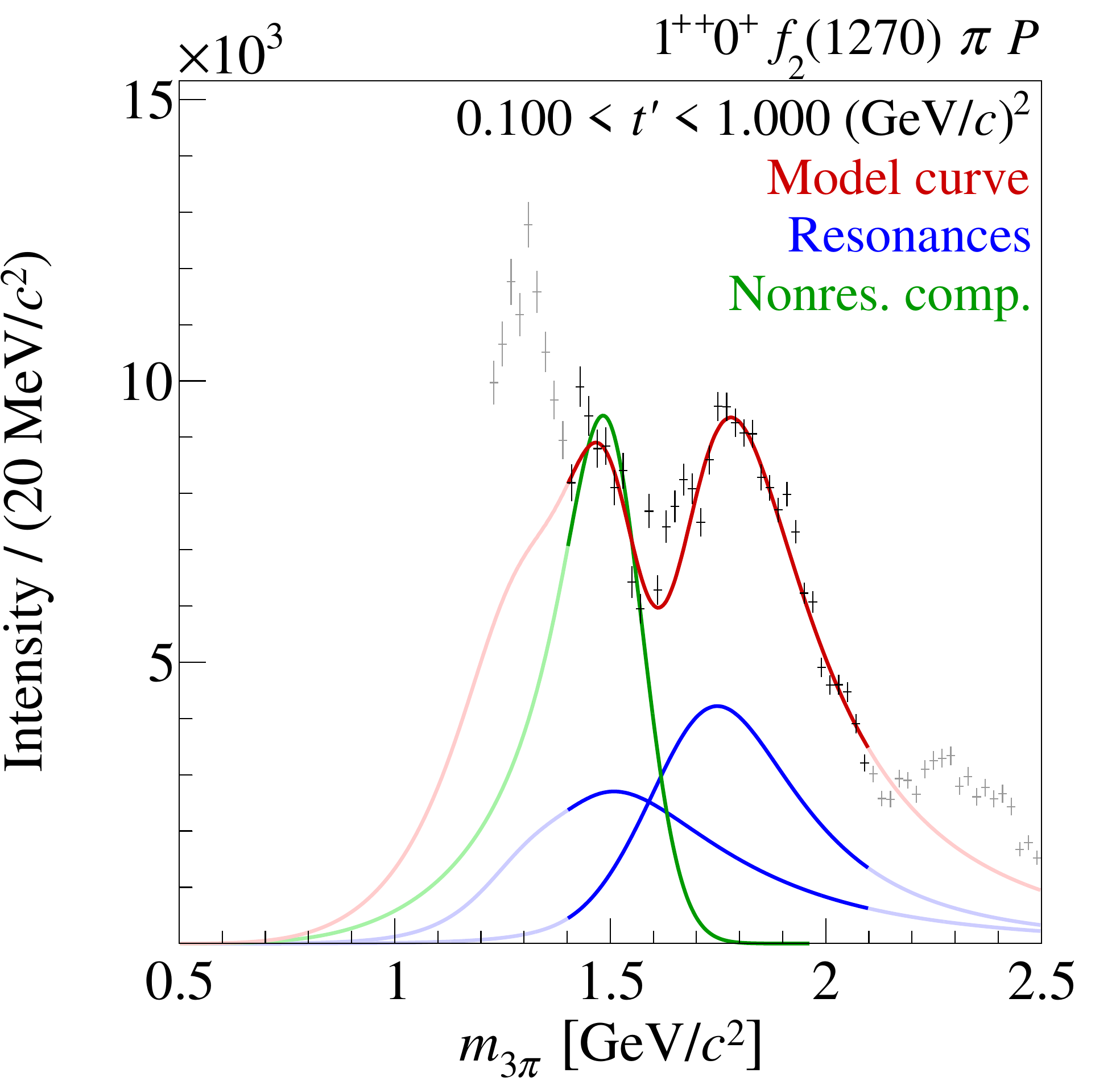}%
			\label{fit:res:1pp:f2}%
		}
		\subfloat[]{%
			\includegraphics[width=\threePlotWidth]{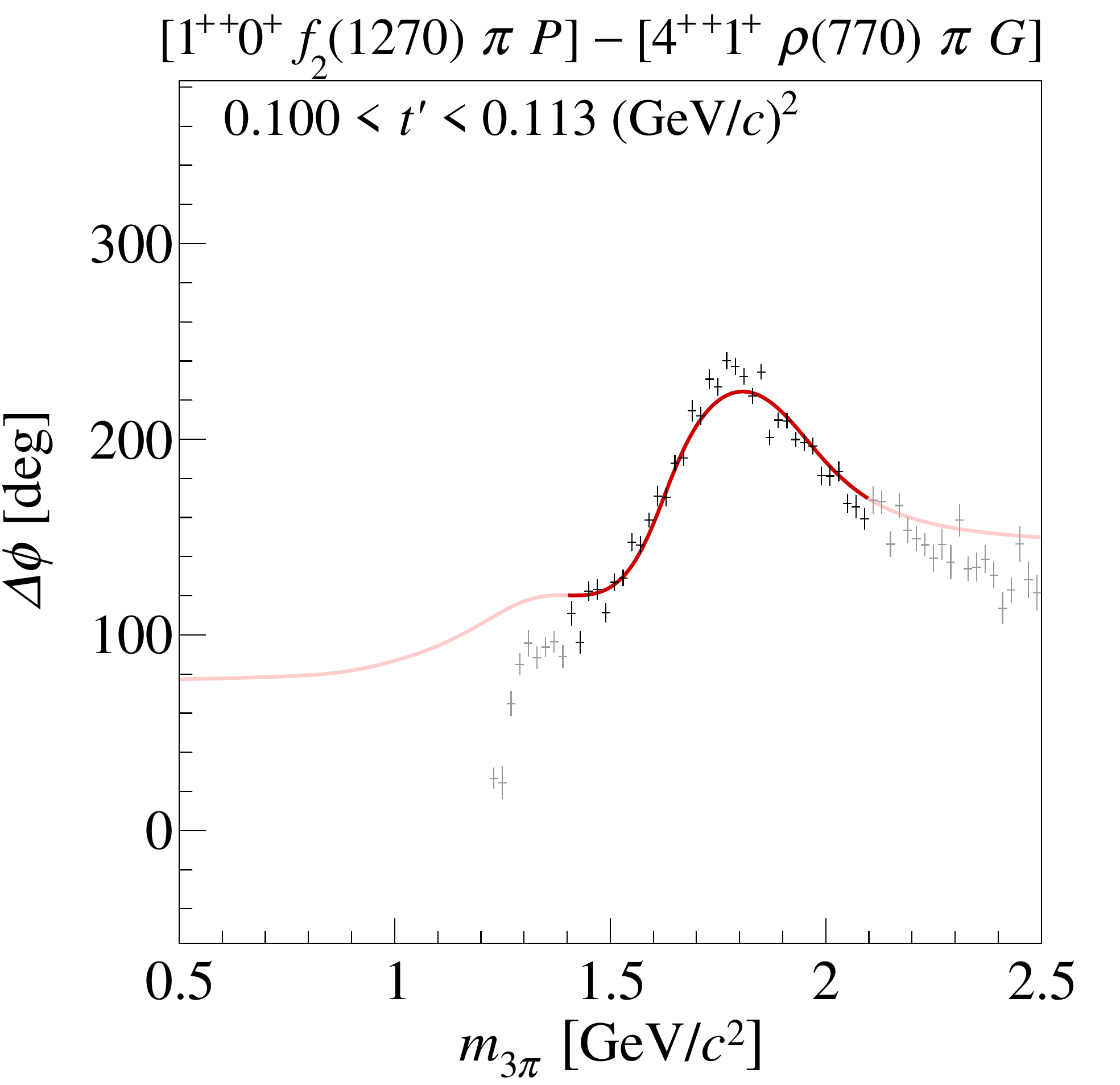}%
			\label{fit:res:1pp:f2phase}%
		}%
		\caption{\tpr-summed intensity distributions of (a) the \Wave 1++0+\Prho\Ppi S and (b) the \Wave 1++0+ \PfTwo \Ppi P wave. (c) shows the relative phase of the \Wave 1++0+ \PfTwo \Ppi P wave with respect to the \Wave 4++1+ \Prho\Ppi G wave in the lowest \tpr bin. Same color code as in \cref{fit:res:4pp} is used.}
		\label{fit:res:1pp}
	\end{figure}
	\begin{figure}[tbp]
		\centering
		\subfloat[]{%
			\includegraphics[width=\twoPlotWidth]{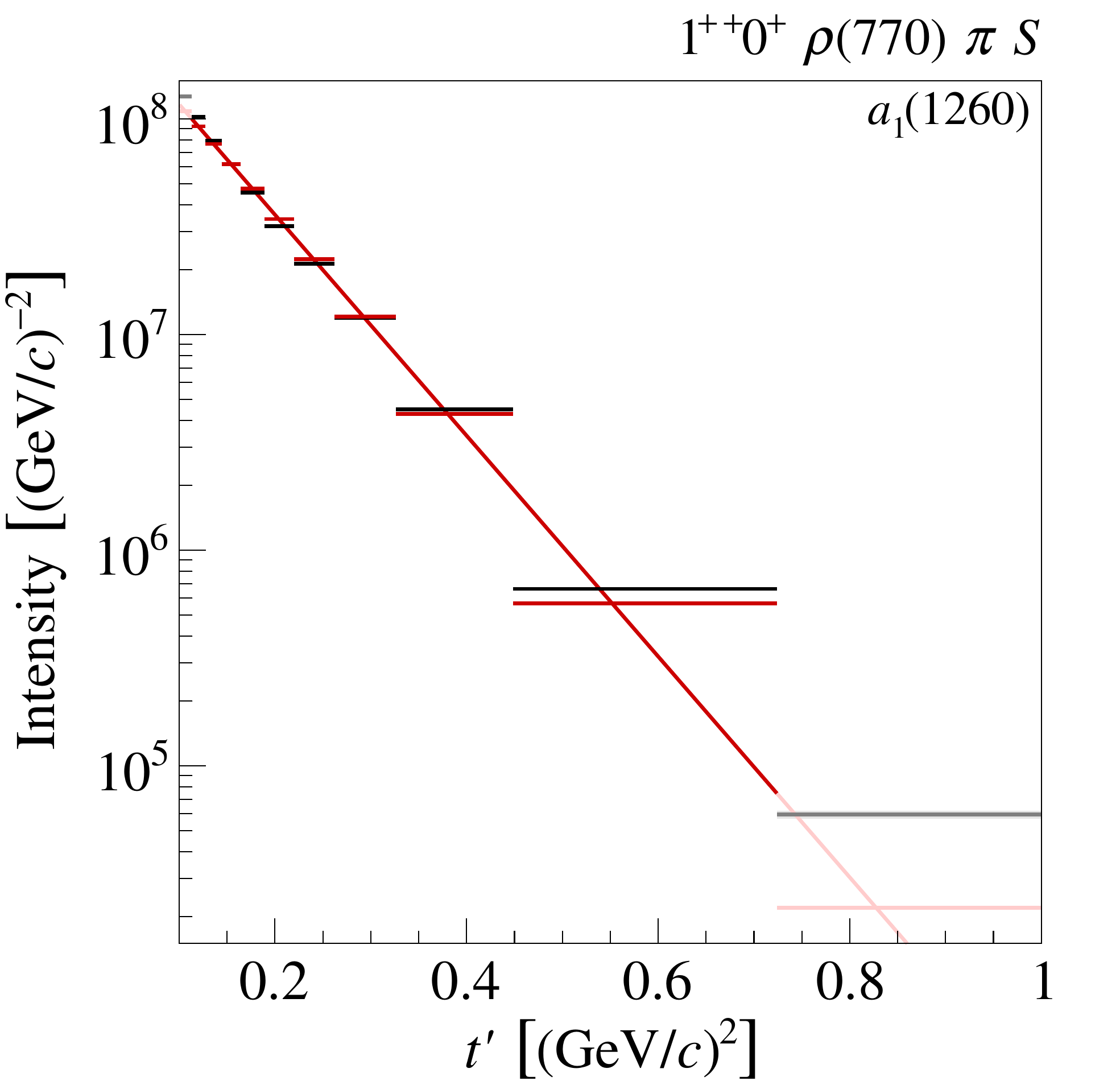}%
			\label{fit:res:1ppt:a1-1260}%
		}%
		\subfloat[]{%
			\includegraphics[width=\twoPlotWidth]{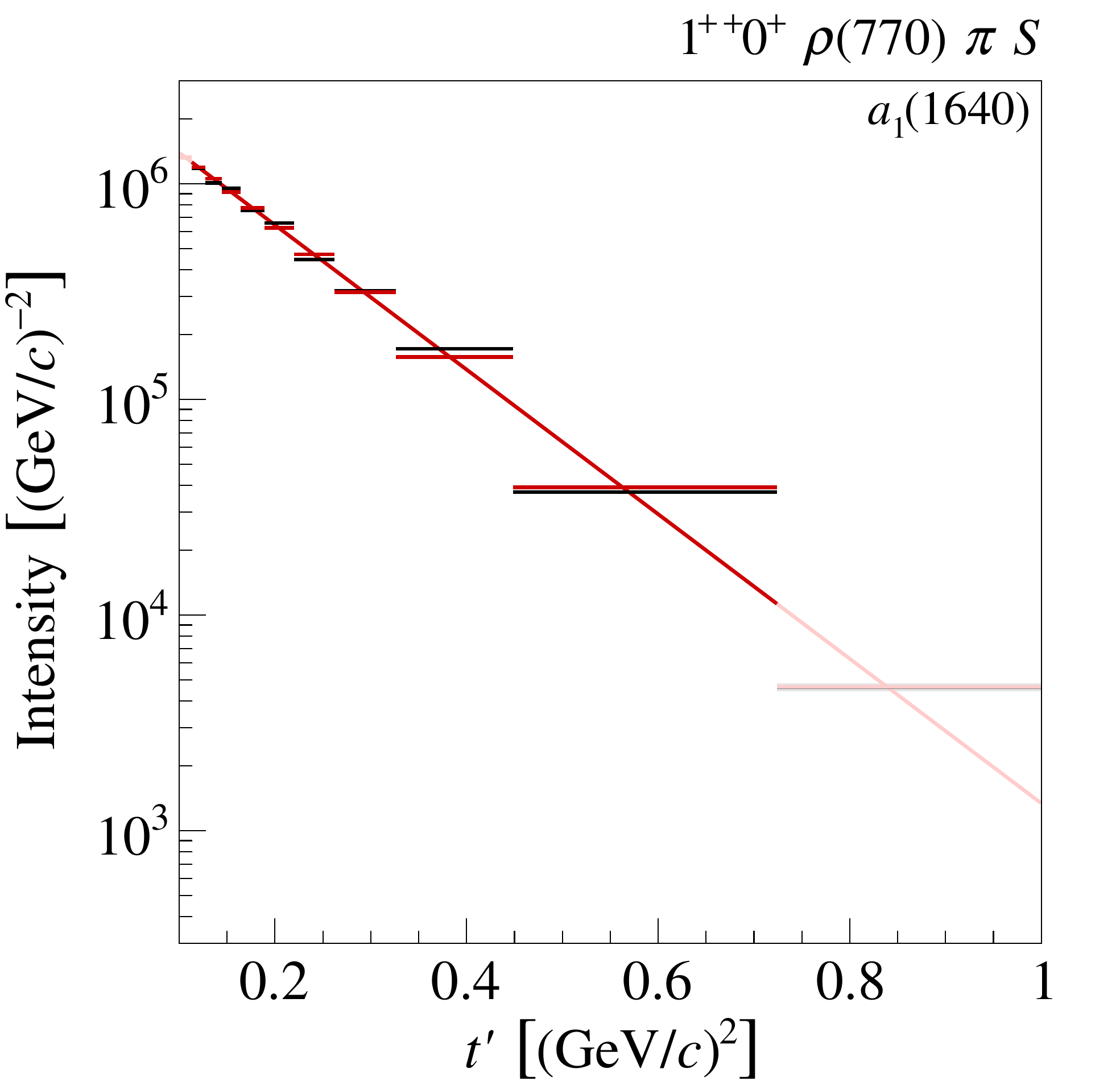}%
			\label{fit:res:1ppt:a1-1640}%
		}%
		\caption{\tpr spectra of (a) the \PaOne and (b) the \PaOnePr in the \Wave 1++0+\Prho\Ppi S wave. Same color code as for the non-resonant components in \cref{fit:res:4ppt} is used.}
		\label{fit:res:1ppt}
	\end{figure}
	The \Wave 1++0+ \Prho\Ppi S wave contributes about \SI{30}{\percent} to the total intensity and  is hence the dominant signal in our data. It exhibits a broad peak in the intensity distribution at about \SI{1.2}{\GeVcc} (see \cref{fit:res:1pp:rho}). This broad peak is described by the \PaOne component and the non-resonant component, which have approximately similar intensities.
	The resonance model cannot describe well the details of the intensity spectrum of the \Wave 1++0+ \Prho\Ppi S wave within the extremely small statistical uncertainties. This is mainly due to the large contribution of the non-resonant component, in combination with our lack of accurate knowledge about the non-resonant shape. This also leads to large systematic uncertainties of the measured \PaOne mass and width of $1299\,^{+12}_{-28}\,\si{\MeVcc}$ and $380\pm 80\,\si{\MeVcc}$, respectively, which are in agreement with the PDG estimates~\cite{Patrignani:2016xqp}.
	
	We observe evidence for a potential \PaOnePr in the high-mass tail of the \PaOne, visible as a shoulder at about \SI{1.8}{\GeVcc} in the intensity spectrum of the \Wave 1++0+\Prho\Ppi S wave. The strongest evidence for the \PaOnePr is observed in the \Wave 1++0+ \PfTwo \Ppi P wave. It shows a clear peak at about \SI{1.8}{\GeVcc} (see \cref{fit:res:1pp:f2}) that is associated with a rising phase motion in this mass region (see \cref{fit:res:1pp:f2phase}). Both features are reproduced well by the resonance model.
	Our estimate for the \PaOnePr parameters are $m_0 = 1700\,^{+35}_{-130}\,\si{\MeVcc}$ and $\Gamma_0 = 510\,^{+170}_{-90}\,\si\MeVcc$. The PDG lists the \PaOnePr as ``omitted from summary table''~\cite{Patrignani:2016xqp}. Our mass value is in agreement the world average, but our estimate for the width is \SI{260}{\MeVcc} larger than the world average. This might be due to the disagreement between model and data in the \PaOne mass region or it might be a consequence of not including any higher-lying \PaOne* states in the fit model. However, in the analyzed \onePP waves, we do not observer clear evidence for further \PaOne* states.
	
	The \tpr spectra of the \PaOne and the \PaOnePr (see \cref{fit:res:1ppt}) are in good agreement with the exponential model in \cref{eq:rm:transition-amplitude}. The slope parameter of the \PaOne of $b = 11.8^{+0.9}_{-4.2}\,\si{\perGeVcsq}$ is unusually large. The \PaOne shows the steepest \tpr spectrum among all resonances included in the resonance-model fit. However, it shows a considerable systematic uncertainty towards smaller slope values. The \PaOne slope value agrees within uncertainties with the slope value of the non-resonant term. This might indicate that the fit is not able to completely separate the \PaOne from the non-resonant part.
	The slope parameter of the \PaOnePr of $b \approx \SI{8}{\perGeVcsq}$ is smaller than the one of the ground state \PaOne and in the range we typically observe for resonance components.

	\subsection{$\bm{\JPC = 2^{-+}}$ resonances}
	\begin{figure}[tbp]
		\centering
		\subfloat[]{%
			\includegraphics[width=\threePlotWidth]{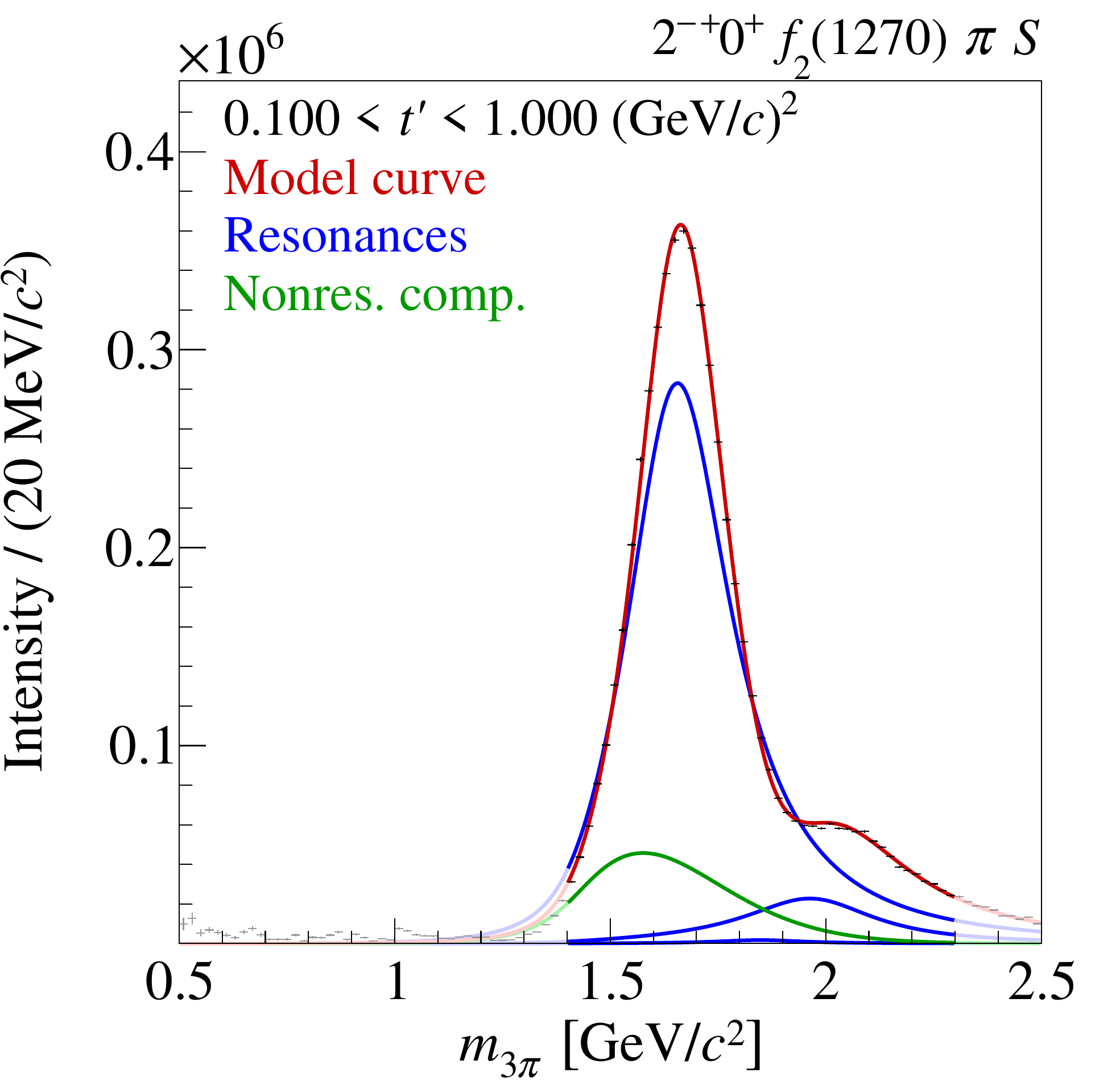}%
			\label{fit:res:2mp:f2S0}%
		}%
		\subfloat[]{%
			\includegraphics[width=\threePlotWidth]{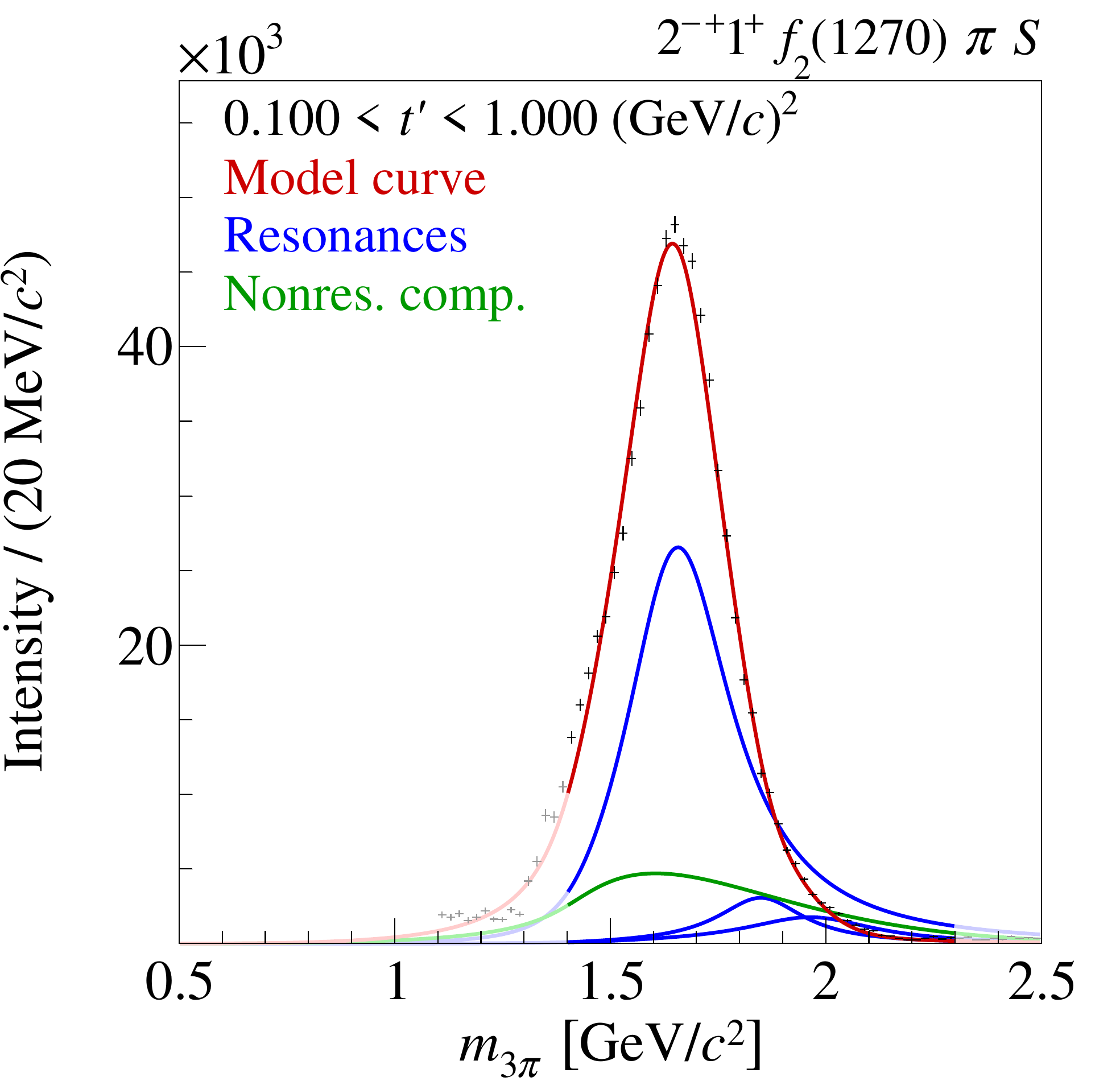}%
			\label{fit:res:2mp:f2S1}%
		}%
		\subfloat[]{%
			\includegraphics[width=\threePlotWidth]{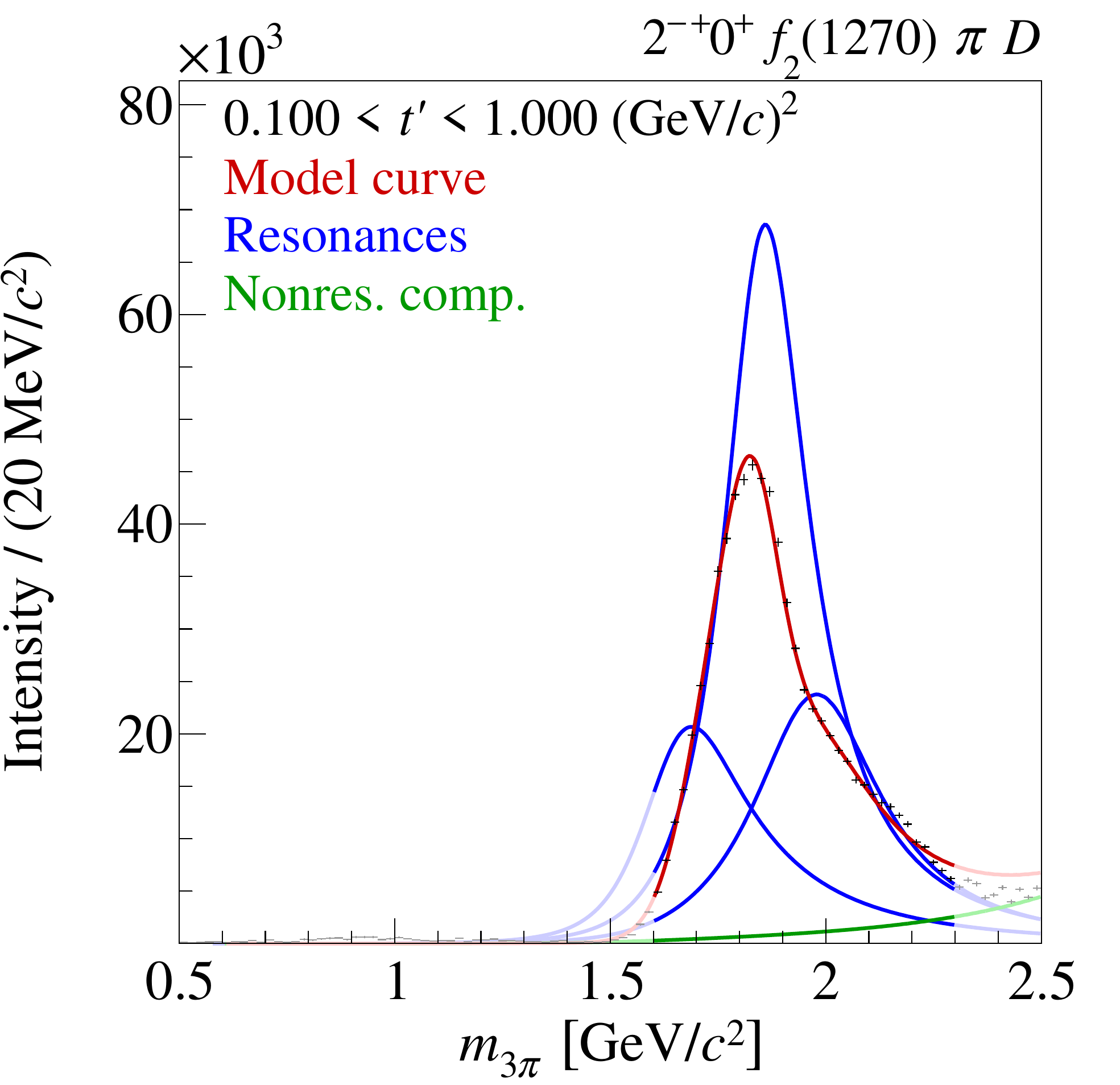}%
			\label{fit:res:2mp:f2D}%
		}\linebreak%
		\subfloat[]{%
			\includegraphics[width=\threePlotWidth]{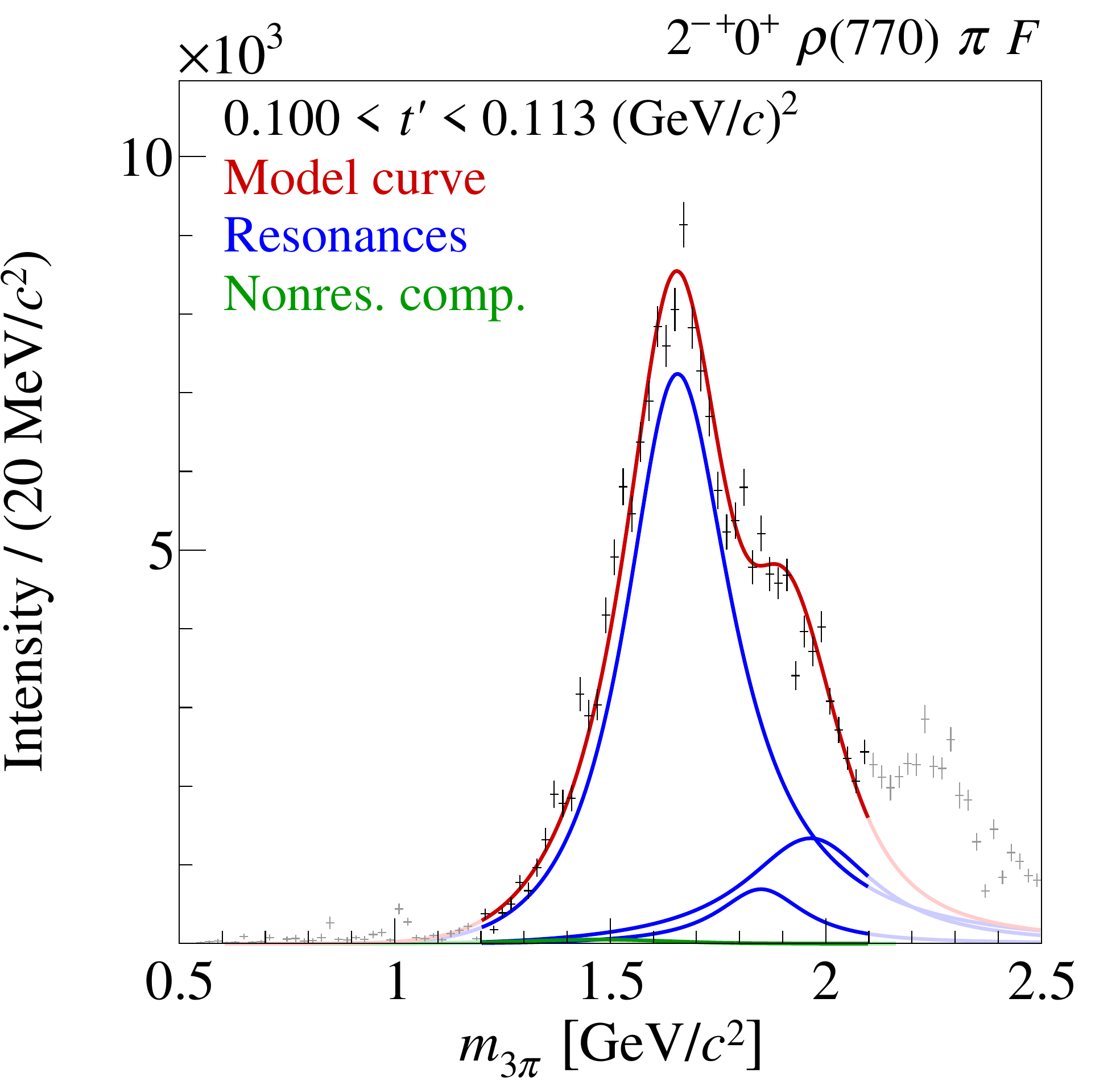}%
			\label{fit:res:2mp:rholowt}%
		}
		\subfloat[]{%
			\includegraphics[width=\threePlotWidth]{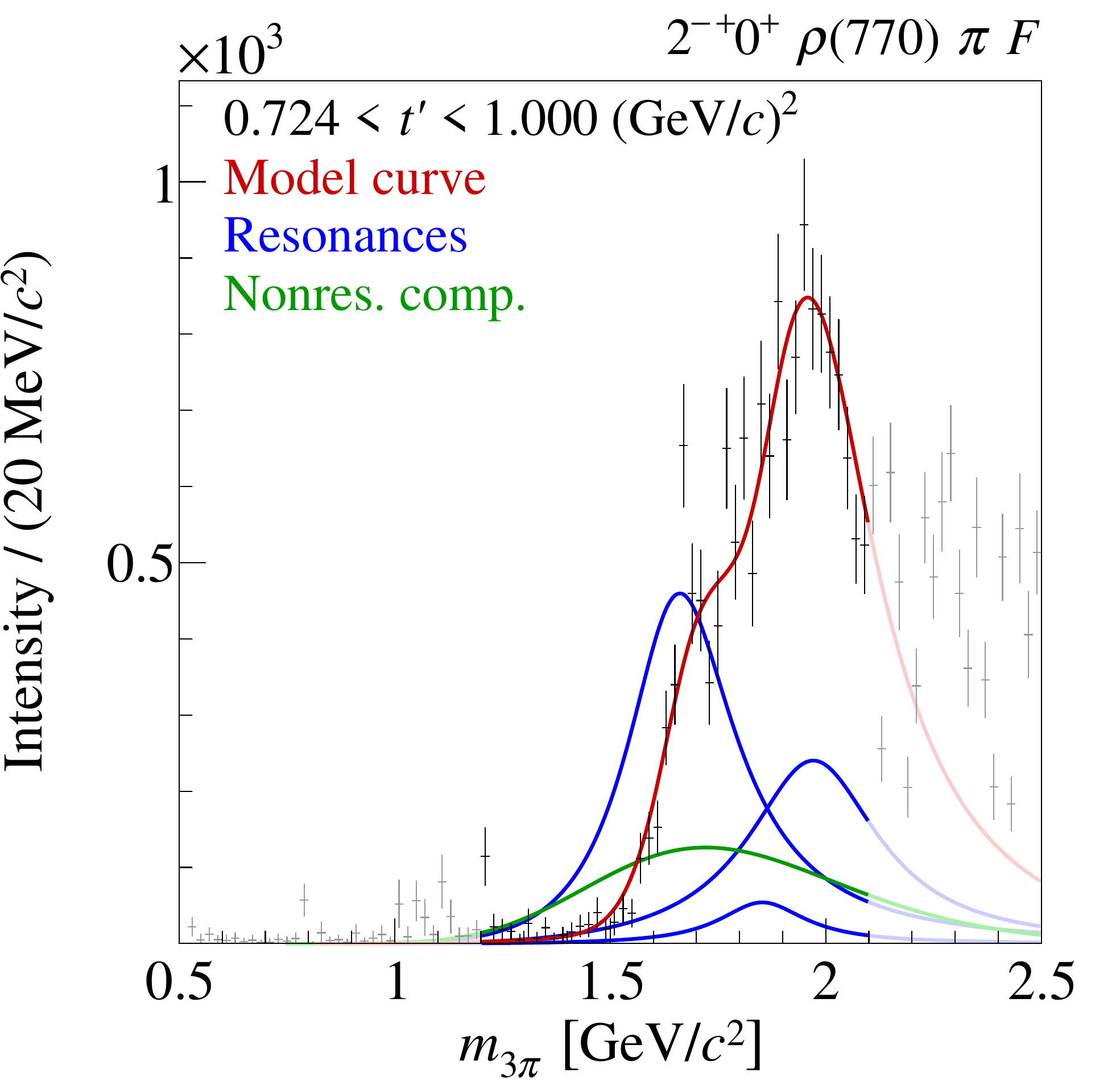}%
			\label{fit:res:2mp:rhohight}%
		}%
		\caption{\tpr-summed intensity distributions of (a) the \Wave 2-+0+\PfTwo\Ppi S, (b) the \Wave 2-+1+\PfTwo\Ppi S, and (c) the \Wave 2-+0+\PfTwo\Ppi D waves. (d) and (e) show the \Wave 2-+0+\Prho\Ppi F partial-wave intensity distribution in the lowest and highest \tpr bin, respectively. Same color code as in \cref{fit:res:4pp} is used.}
		\label{fit:res:2mp}
	\end{figure}
	\begin{figure}[tbp]
		\subfloat[]{%
			\includegraphics[width=\threePlotWidth]{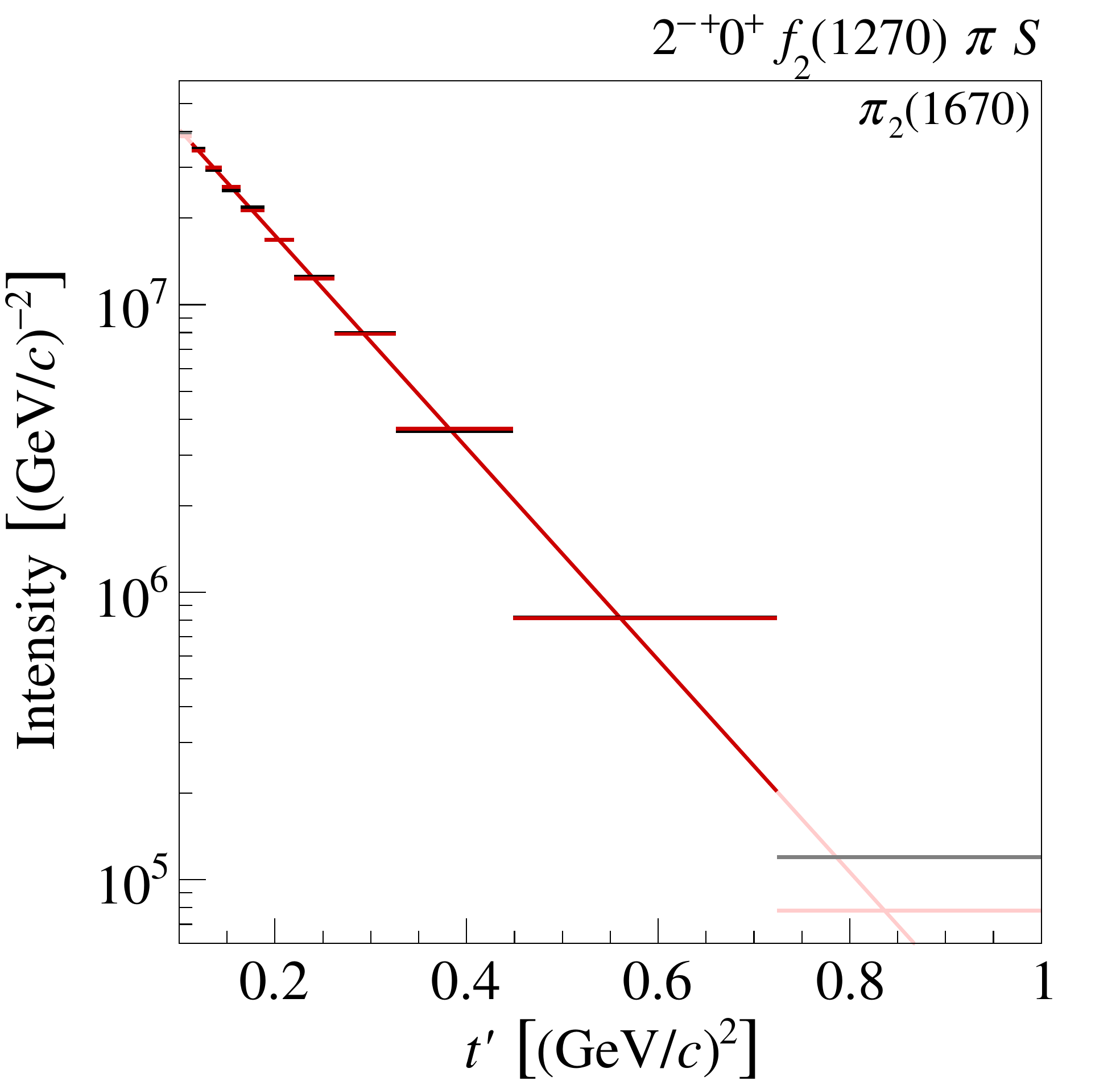}%
			\label{fit:res:2mpt:pi2-1670}%
		}%
		\subfloat[]{%
			\includegraphics[width=\threePlotWidth]{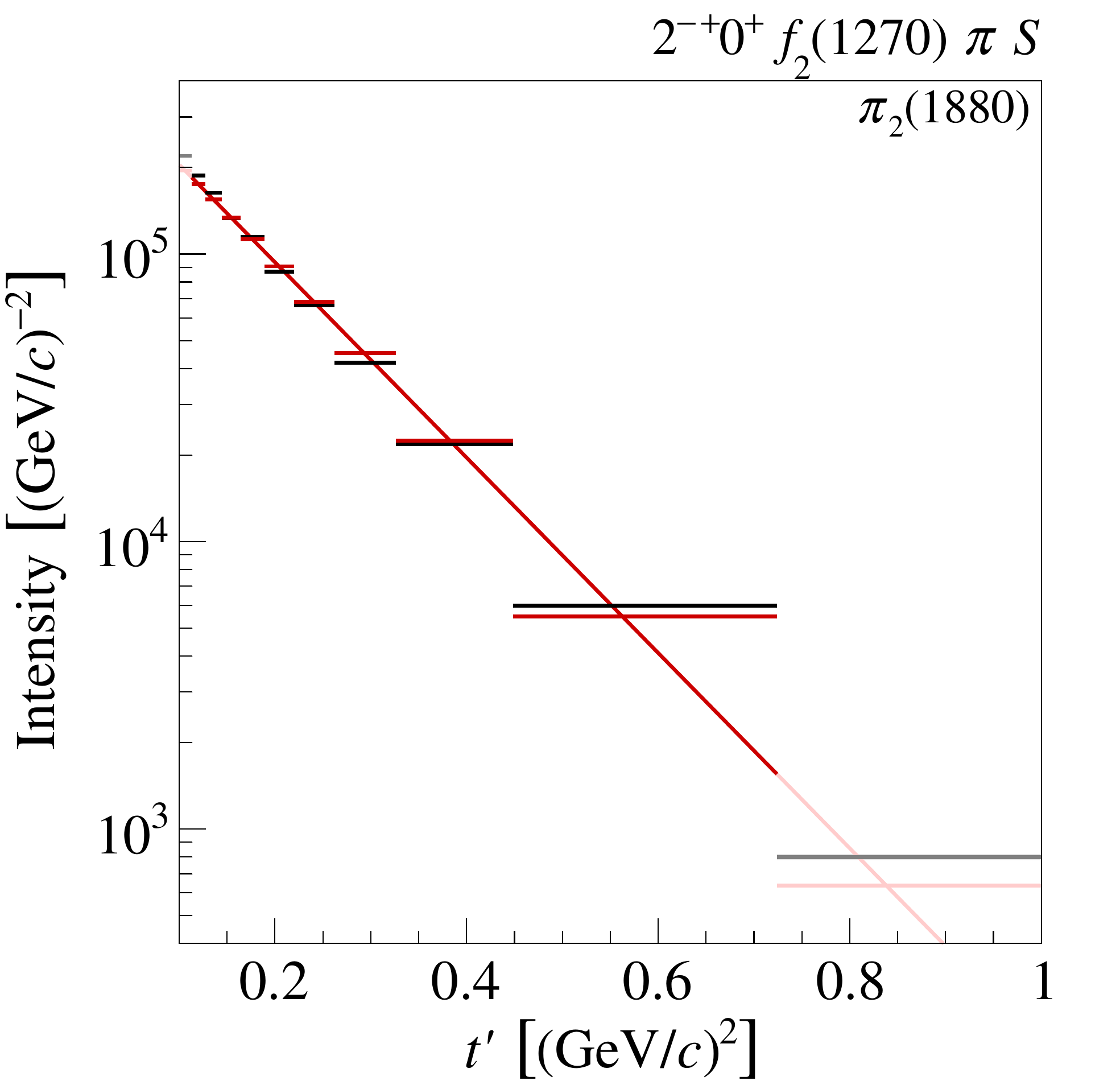}%
			\label{fit:res:2mpt:pi2-1880}%
		}%
		\subfloat[]{%
			\includegraphics[width=\threePlotWidth]{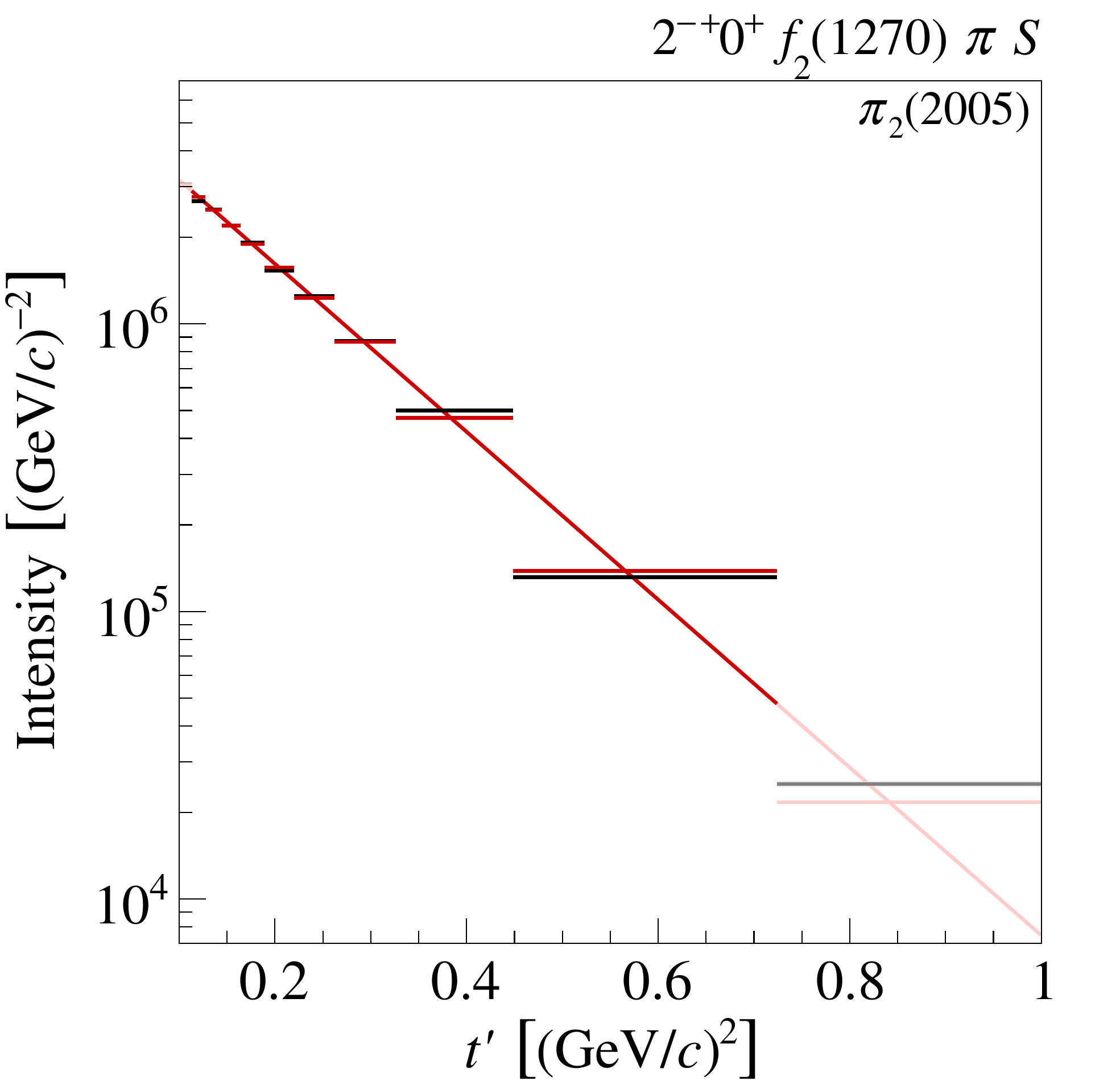}%
			\label{fit:res:2mpt:pi2-2005}%
		}%
		\caption{\tpr spectra of (a) the \PpiTwo, (b) the \PpiTwoPr, and (c) the \PpiTwoPrPr in the \Wave 2-+0+\PfTwo\Ppi S wave. Same color code as for the non-resonant components in \cref{fit:res:4ppt} is used.}
		\label{fit:res:2mpt}
	\end{figure}
	\begin{figure}[tbp]
		\subfloat[]{%
			\includegraphics[width=\threePlotWidth]{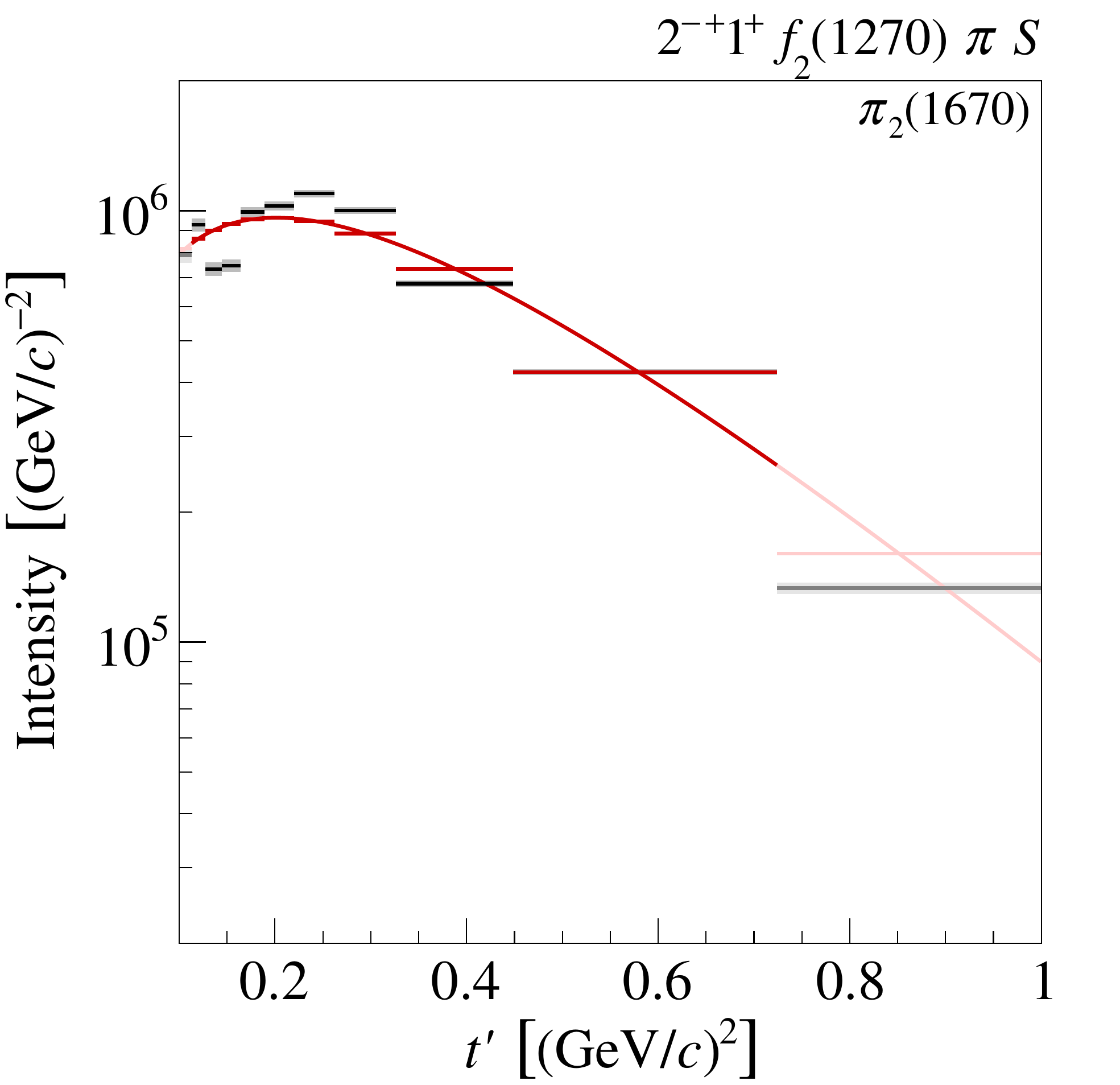}%
			\label{fit:res:2mp1t:pi2-1670}%
		}%
		\subfloat[]{%
			\includegraphics[width=\threePlotWidth]{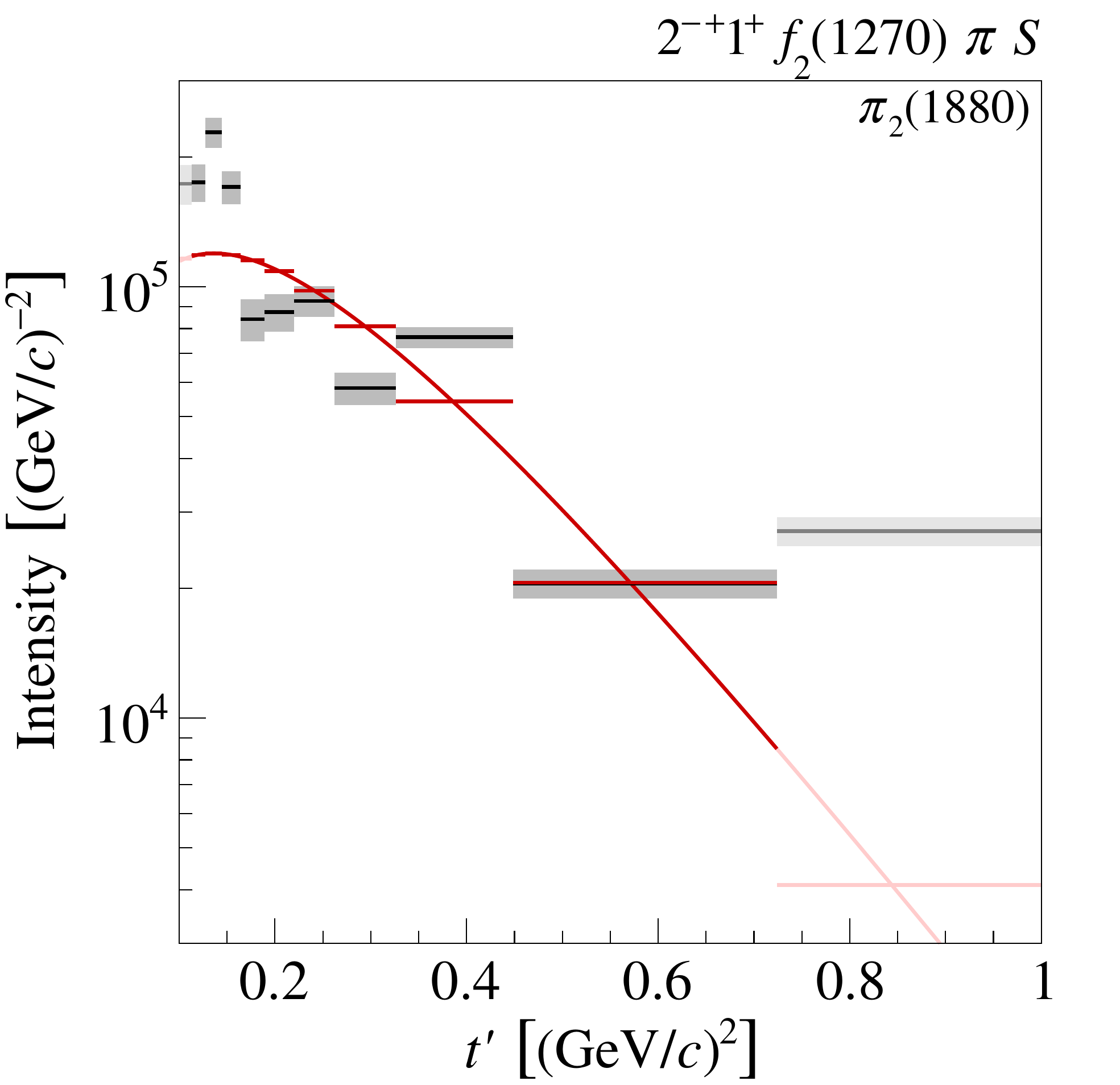}%
			\label{fit:res:2mp1t:pi2-1880}%
		}%
		\subfloat[]{%
			\includegraphics[width=\threePlotWidth]{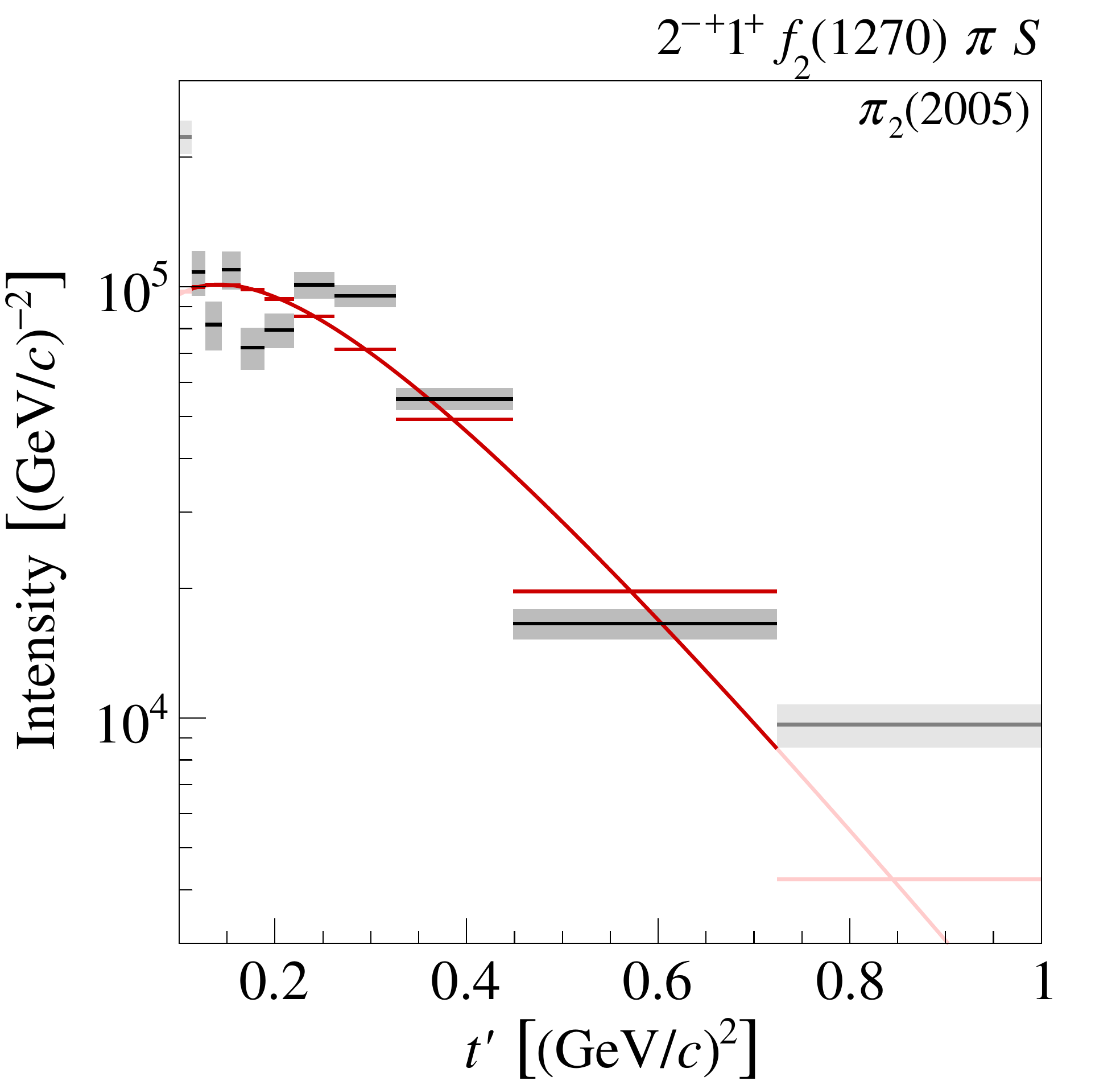}%
			\label{fit:res:2mp1t:pi2-2005}%
		}%
		\caption{\tpr spectra of (a) the \PpiTwo, (b) the \PpiTwoPr, and (c) the \PpiTwoPrPr in the \Wave 2-+1+\PfTwo\Ppi S wave. Same color code as for the non-resonant components in \cref{fit:res:4ppt} is used.}
		\label{fit:res:2mp1t}
	\end{figure}
	We include four partial waves with $\JPC=\twoMP$ in the resonance-model fit, each of which is parameterized by three \PpiTwo* resonance components and a non-resonant component.
	Of the four waves, the \Wave 2-+0+\PfTwo\Ppi S wave has the largest intensity. It exhibits a striking peak at about \SI{1.65}{\GeVcc} (see \cref{fit:res:2mp:f2S0}). This peak is reproduced well by a dominant contribution of the \PpiTwo component. We also include the \Wave 2-+{1}+\PfTwo\Ppi S wave with spin-projection $M=1$ (see \cref{fit:res:2mp:f2S1}), which shows similar features as the $M=0$ wave.
	
	The \Wave 2-+0+\PfTwo\Ppi D wave exhibits a striking peak as well. However, not at \SI{1.65}{\GeVcc}, but at about \SI{1.8}{\GeVcc} (see \cref{fit:res:2mp:f2D}). The peak is described mainly by the \PpiTwoPr component. The low- and high-mass tails of the peak are described as an interference effect among the \PpiTwoPr, the \PpiTwo, and the \PpiTwoPrPr components.
	
	The \Wave 2-+0+\Prho\Ppi F wave shows the strongest evidence for the \PpiTwoPrPr. 
	In the low-\tpr region (see \cref{fit:res:2mp:rholowt}), the intensity spectrum is dominated by a clear peak at about \SI{1.65}{\GeVcc}, mainly described by the \PpiTwo component, with a small peak at about \SI{2}{\GeVcc} in its high-mass shoulder.
	In the high-\tpr region (see \cref{fit:res:2mp:rhohight}), the higher-lying peak dominates the intensity spectrum and is mainly described by the \PpiTwoPrPr, while the \PpiTwo is visible only as a low-mass shoulder.
	
	\Cref{fit:res:2mpt} shows the \tpr spectra of the three \PpiTwo* resonances in the \Wave 2-+0+\PfTwo\Ppi S wave. They are in good agreement with the exponential model in \cref{eq:method:tspectrum}. The extracted slope parameters of $b = 8.5^{+0.9}_{-0.5}\,\si\perGeVcsq$ for the \PpiTwo, $b = 7.8^{+0.5}_{-0.9}\,\si\perGeVcsq$ for the \PpiTwoPr, and $b = 6.7^{+0.5}_{-1.3}\,\si\perGeVcsq$ for the \PpiTwoPrPr lie within the range we typically observe for resonances.
	We observe a similar pattern as for the ground and excited \PaOne* states, that the slope parameters decrease for increasing resonance masses.
	The \tpr spectra of the resonances in the other two \twoMP partial waves with $M=0$ are constrained by \cref{eq:method:branching} to have the same shapes as the ones in of the \Wave 2-+0+\PfTwo\Ppi S wave. However, in the \Wave 2-+1+\PfTwo\Ppi S wave, the \tpr dependence of the resonance components is independent from the ones in the $M=0$ waves. \Cref{fit:res:2mp1t}  shows the \tpr spectra of the components in the \Wave 2-+1+\PfTwo\Ppi S wave.
	In general, the \tpr spectra of the components in this wave are extracted less reliably because: (i) they are constrained by only one wave, while the $M=0$ spectra are constrained by three waves and (ii) the \Decay\PfTwo\Ppi S wave with $M=1$ is small compared to the one with $M=0$.
	The \tpr spectra for the $M=1$ wave are in rough agreement with the exponential model in \cref{eq:method:tspectrum}. In particular, they exhibit a drop in intensity towards small values of \tpr as predicted by the model (see \cref{sec:method:rmf}).
	The slope parameters for the \PpiTwoPr and the \PpiTwoPrPr in the $M=1$ wave are consistent with the ones in the $M=0$ waves. However, for the \PpiTwo, we estimate a slope parameter of $b \approx \SI{5.0}{\perGeVcsq}$ for the $M=1$ wave, which is significantly smaller than for the $M=0$ waves. This effect is not understood, but it is consistent with the shallower \tpr spectrum of the total intensity in the $M=1$ wave.
	
	\section{$\mathbf{\tpr}$ dependence of relative phases between resonances}
	\begin{figure}[bth]
		\begin{center}
			\includegraphics[width=0.7\linewidth]{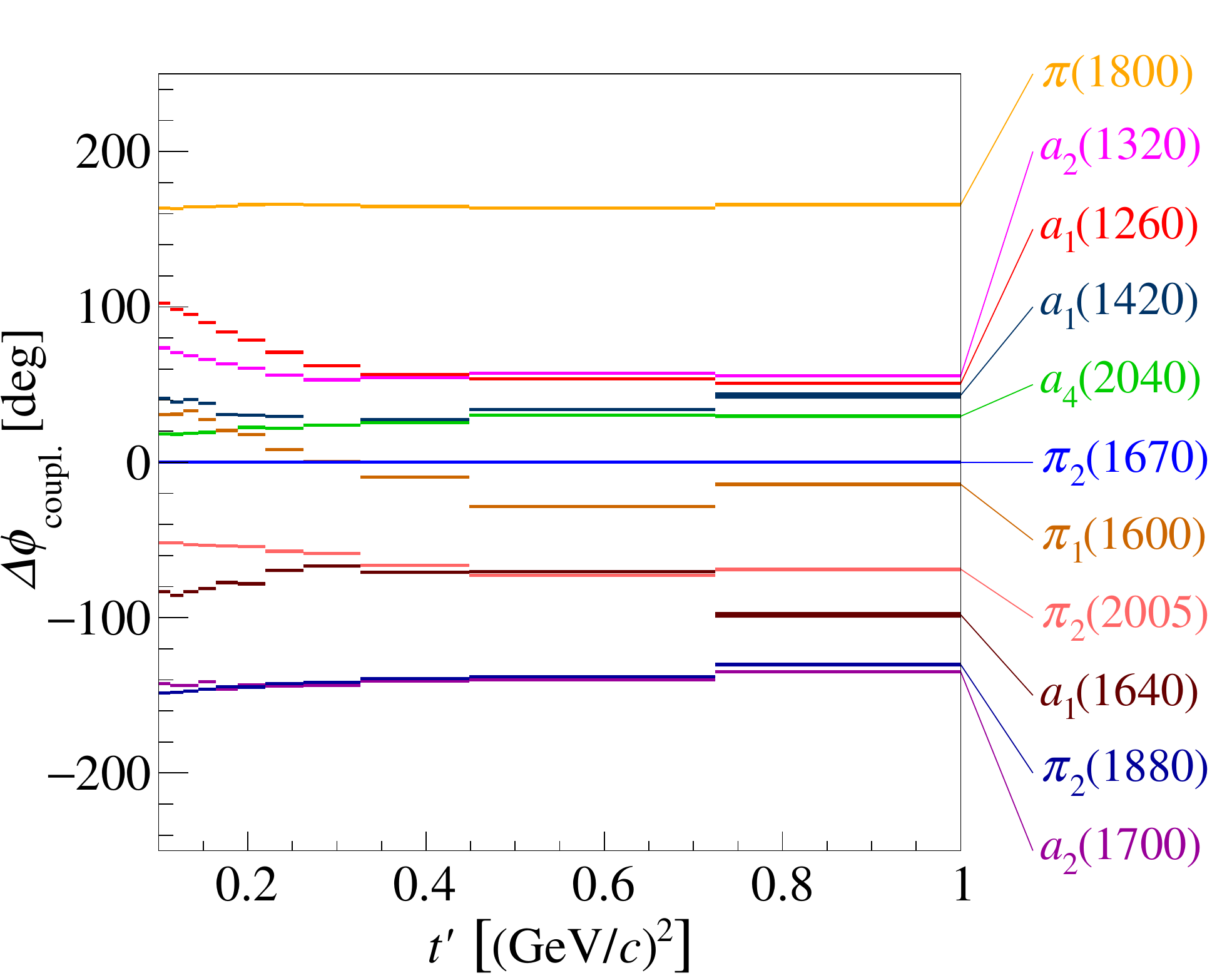}
		\end{center}
		\caption{\tpr dependence of the relative phases
			$\Delta \phi_\text{coupl.}$ of the coupling amplitudes of the 11~resonance components in the 
			fit model with respect to the \PpiTwo.  The coupling phases are shown for the 
			dominant wave of the respective \JPC sector:
			\wave{0}{-+}{0}{+}{\PfZero}{S}, \wave{1}{++}{0}{+}{\Prho}{S},
			\wave{1}{-+}{1}{+}{\Prho}{P}, \wave{2}{++}{1}{+}{\Prho}{D},
			\wave{2}{-+}{0}{+}{\PfTwo}{S}, and \wave{4}{++}{1}{+}{\Prho}{G}.
			The only exception is the \PaOne[1420], which appears only in the 
			\wave{1}{++}{0}{+}{\PfZero}{P} wave.  The width of the horizontal
			lines represents the statistical uncertainty. The systematic
			uncertainty is not shown.
		}
		\label{fig:tphase:all}
	\end{figure}
	\begin{figure}[tbp]
		\centering
		\subfloat[]{%
			\includegraphics[width=\twoPlotWidth]{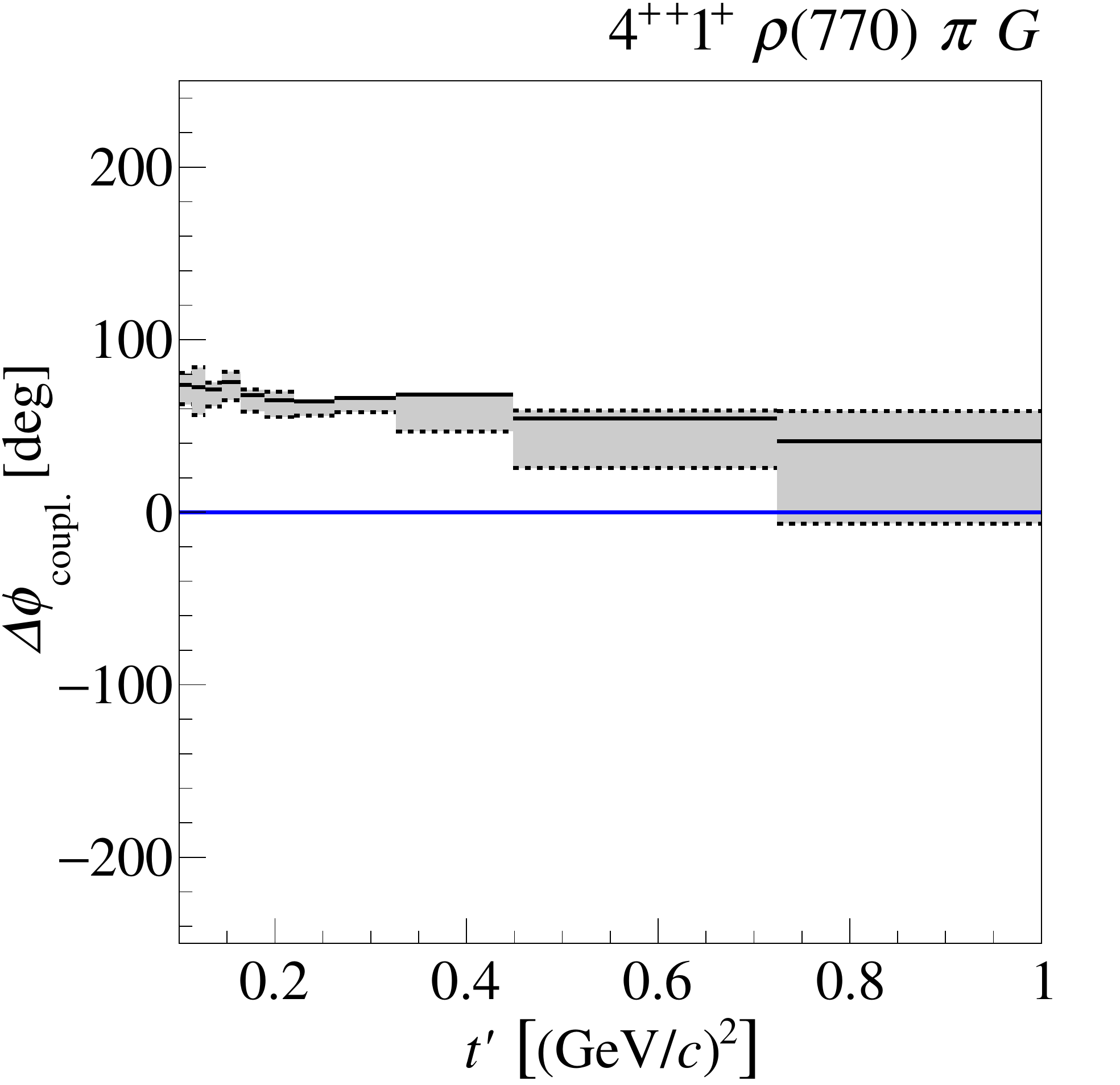}%
			\label{fig:tphase:a4:rho}%
		}%
		\subfloat[]{%
			\includegraphics[width=\twoPlotWidth]{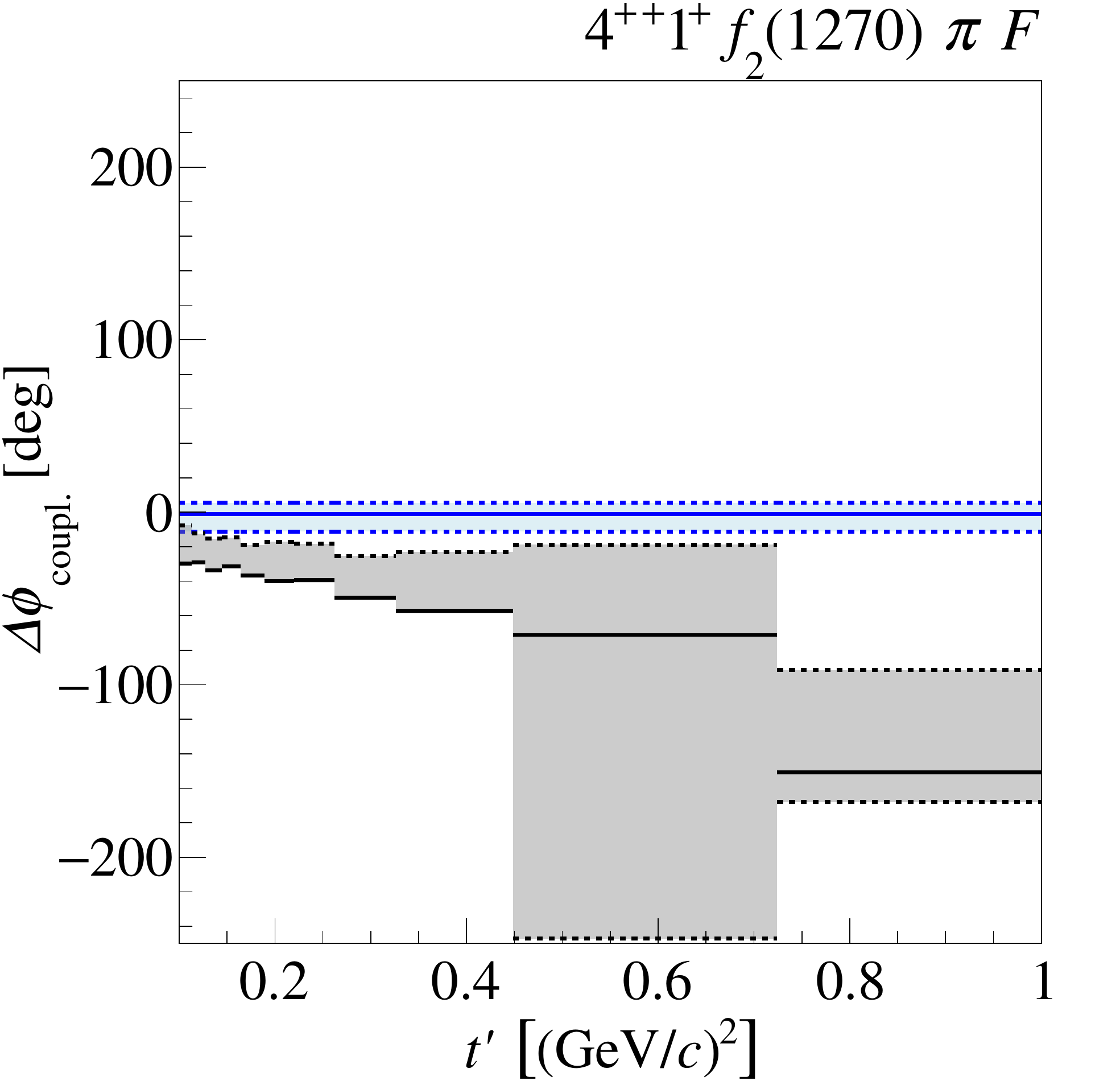}%
			\label{fig:tphase:a4:f2}%
		}%
		\caption{\tpr dependence of the coupling phase of the wave
			components in (a)~the
			\wave{4}{++}{1}{+}{\Prho}{G} and (b)~the
			\wave{4}{++}{1}{+}{\PfTwo}{F} wave.  The coupling phases of the
			\PaFour (blue lines) and of the non-resonant component (black lines)
			are shown relative to the \PaFour in the
			\wave{4}{++}{1}{+}{\Prho}{G} wave. For each wave component, the
			magnitude of the effects observed in the systematic studies is illustrated qualitatively by two sets
			of dashed lines with shaded area in between (see ref.~\cite{Adolph2018} for details).}
		\label{fig:tphase:a4}
	\end{figure}
	
	In addition to the \tpr dependence of the intensity of a wave component $j$ in wave $a$, i.e. its \tpr spectrum, we can extract also the \tpr dependence of the relative phase of the coupling amplitude of wave component $j$ in wave $a$ with respect to the wave component $k$ in wave $b$. This is called the \textit{coupling phase}:
	\begin{align}
		\Delta \phi_\mathrm{coupl.}^{j, a; k, b}(\tpr) \equiv \arg\left[\mathcal{C}^j_a(\tpr)\,\mathcal{C}^{k*}_b(\tpr)\right]
	\end{align}
	
	\Cref{fig:tphase:all} shows the coupling phases of all 11~resonance components included in the resonance model relative to the \PpiTwo in the \Wave 2-+0+\PfTwo\Ppi S wave. We observe three striking features of the \tpr dependence of the coupling phases of the resonances. First, for $\tpr \gtrsim \SI{0.3}{\GeVcc}$ the phases level off, while for $\tpr \lesssim \SI{0.3}{\GeVcc}$ most of the resonances show a slight change of the coupling phase with respect to the \PpiTwo. Second, ground and excited states show large relative phase offsets in the high-\tpr region, with the exception of the \PaOneFourteenTwenty. The phase of the \PaOneFourteenTwenty is very similar to the one of the ground-state \PaOne, which points to the peculiar nature of this signal (see ref.~\cite{Adolph2018} for details). Third, in the high-\tpr region the coupling phases of ground-state resonances do not deviate by more than $\pm \SI{60}{\degree}$ from the phase of the \PpiTwo. These features are consistent with a common production mechanism for the observed signals.
	
	We can also study the coupling phases of the same state in different decay modes. \Cref{fig:tphase:a4} shows the coupling phases of the \PaFour in the $\Prho\Ppi G$ and $\PfTwo\Ppi F$ decay modes together with the coupling phases of the corresponding non-resonant components in these waves. As the relative coupling phase of the \PaFour in the $\Prho\Ppi G$ decay is $\SI{0}{\degree}$ by definition, its coupling phase in the $\PfTwo\Ppi F$ decay has to be a constant as both are related by \cref{eq:method:branching}. However, the model allows an arbitrary phase offset between the two decay modes. We estimate this phase offset to be close to $\SI{0}{\degree}$ with small systematic uncertainties.
	The non-resonant components in the two waves show a slight variation of their coupling phase with \tpr with respect to the \PaFour resonance. This feature is also observed for the non-resonant components in other partial waves.

	\section{Conclusion}
	We have performed the so far largest resonance-model fit that simultaneously and consistently describes 14 partial waves including their mutual interferences by a single model.
	The huge amount of information condensed in this fit, in combination with the information from the our novel \tpr-resolved analysis approach allows us to study ground  as well as excited states.
	It also allows us to study in detail the \tpr dependence of the intensity and coupling phase of these resonances.
	We observe that for most partial waves, the \tpr spectra of the non-resonant components fall steeper with \tpr as compared to the resonances. This helps to separate resonant from non-resonant contributions.
	We also observe that for most of the resonances, the \tpr spectra become shallower with increasing resonance mass. This is consistent with the observation that the slope of the overall \tpr spectrum of the data becomes shallower for higher \mThreePi masses (see ref.~\cite{Adolph2015}). Furthermore, the different \tpr dependences of the ground-state and the excited resonances helps to better separate them. An example is the strong evolution of the intensity spectrum of the \Wave 2-+0+\Prho\Ppi F wave with \tpr (see \cref{fit:res:2mp:rholowt,fit:res:2mp:rhohight}).
	Our approach also allows to study the \tpr dependence of the relative phases of the resonances. The pattern we observe is consistent with a common production mechanism.
	To reduce the systematic uncertainties and to further clarify the existence of higher excited states, models that are based on the principles of unitarity and analyticity are mandatory, especially in waves where the non-resonant contributions are important. This is the topic of future research.
	
	\printbibliography[heading=bibintoc]
	
\end{document}